\PassOptionsToPackage{table, prologue}{xcolor}

\documentclass[sigconf, nonacm]{acmart}

\AtBeginDocument{%
  \providecommand\BibTeX{{%
    \normalfont B\kern-0.5em{\scshape i\kern-0.25em b}\kern-0.8em\TeX}}}

\usepackage{algorithm}
\usepackage{algorithmic}
\usepackage{amsmath}
\usepackage{booktabs}
\usepackage{soul} 
\usepackage{color, xcolor}
\usepackage{xspace}
\usepackage{diagbox}
\usepackage{subcaption}
\usepackage{multirow}
\usepackage{siunitx}
\usepackage{wrapfig}
\usepackage{graphicx}

\usepackage{xcolor}
\usepackage{makecell}

\usepackage{url}

\usepackage{hyperref}
\usepackage{enumitem}
\usepackage{makecell}
\usepackage{tikz}
\usetikzlibrary{arrows.meta,
                chains,
                positioning,
                shapes.geometric
                }
                
\usepackage{newfloat}
\usepackage{listings}

\begin{document}

\title{Global News Synchrony and Diversity During the Start of the COVID-19 Pandemic}


\author{Xi Chen}
\affiliation{%
  \institution{University of Massachusetts Amherst}
  \city{Amherst}
  \state{Massachusetts}
  \country{USA}
}
\email{xchen4@umass.edu}

\author{Scott A.\ Hale}
\affiliation{%
  \institution{University of Oxford \& Meedan}
  \city{Oxford}
  \country{United Kingdom}}
\email{scott.hale@oii.ox.ac.uk}

\author{David Jurgens}
\affiliation{%
  \institution{University of Michigan}
  \city{Ann Arbor}
  \state{Michigan}
  \country{USA}
}
\email{jurgens@umich.edu}

\author{Mattia Samory}
\affiliation{%
 \institution{Sapienza University of Rome}
 \city{Rome}
 \country{Italy}
 }
\email{mattia.samory@uniroma1.it}

\author{Ethan Zuckerman}
\affiliation{%
  \institution{University of Massachusetts Amherst}
  \city{Amherst}
  \state{Massachusetts}
  \country{USA}}
\email{ethanz@umass.edu}

\author{Przemyslaw A. Grabowicz}
\affiliation{%
  \institution{University of Massachusetts Amherst}
  \city{Amherst}
  \state{Massachusetts}
  \country{USA}}
\email{grabowicz@cs.umass.edu}


\renewcommand{\shortauthors}{Xi Chen et al.}

\begin{abstract}

News coverage profoundly affects how countries and individuals behave in international relations. Yet, we have little empirical evidence of how news coverage varies across countries.
To enable studies of global news coverage, we develop an efficient computational methodology that comprises three components: (i) a transformer model to estimate multilingual news similarity; (ii) a global event identification system that clusters news based on a similarity network of news articles; and (iii) measures of news synchrony across countries and news diversity within a country, based on country-specific distributions of news coverage of the global events. Each component achieves state-of-the art performance, scaling seamlessly to massive datasets of millions of news articles.

We apply the methodology to 60 million news articles published globally between January 1 and June 30, 2020, across 124 countries and 10 languages, detecting 4357 news events. We identify the factors explaining diversity and synchrony of news coverage across countries.
Our study reveals that news media tend to cover a more diverse set of events in countries with larger Internet penetration, more official languages, larger religious diversity, higher economic inequality, and larger populations. Coverage of news events is more synchronized between countries that not only actively participate in commercial and political relations---such as, pairs of countries with high bilateral trade volume, and countries that belong to the NATO military alliance or BRICS group of major emerging economies---but also countries that share certain traits: an official language, high GDP, and high democracy indices.

\end{abstract}






\settopmatter{printacmref=false}

\maketitle

\section{Introduction}
\label{sec:introduciton}

\begin{figure}[t]
    \centering
    \includegraphics[width=0.90\linewidth]{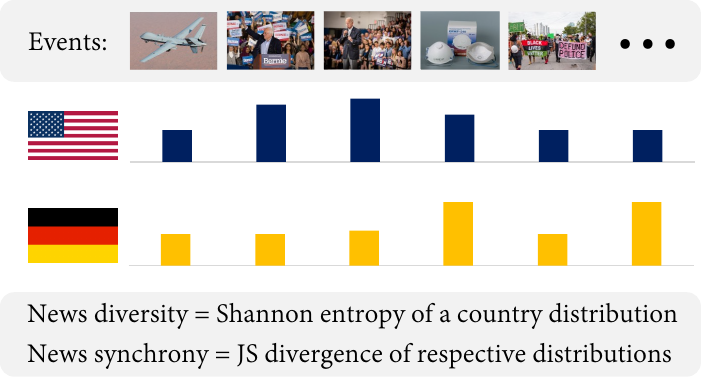}
    \caption{For each country, we measure the distribution of news published in the country across the identified 4357 news events. Based on the distributions we compute news diversity within a country and news synchrony across countries.}
    \label{fig:intro}
\end{figure}

When news media choose to cover a story, they shape the way the public perceives, understands, and discusses current events \cite{mccombs2002news}. By selecting the events that are salient, news media serve a crucial role in shaping the information and political ecosystems. Yet, the selection of news may differ from location to location even at the same moment, and therefore readers over the globe may be offered different images of the present. 

%
Media and communication scholars offer insights on the factors that may lead news media of different countries to cover similar or diverse events. Selection of the news may be driven by the interests of the readers~\cite{paulussen2021news}, but also by larger forces in the information market such as editorial practices and agenda setting \cite{galtung1965structure, zuckerman2003global}. Across borders, several factors including cultural or political affinity, geographic proximity, and economic relations may impact the synchrony of news coverage \cite{kim1996determinants}. 

These insights, however, derive mainly from theoretical frameworks and case studies with limited geographic and event coverage. Empirical studies characterized news coverage within a limited set of countries~\cite{jansen2019drives, boomgaarden2010news}, or when they included more countries, they focused on a limited set of news events~\cite{baum2015filtering, yesilbas2021analysis}. Hardly any study compares coverage of events across multiple countries and languages, necessary to obtain insights into global agenda setting. 

The key challenges to examining the news coverage at a global scale include the following. 
First, traditional data collection and validation based on physical newspapers and questionnaires requires human effort that scales linearly with the amount of data and the number of languages. 
These practical considerations severely limit data size and linguistic coverage of traditional studies.
Second, it is not clear how to identify global news events, which are necessary to measure news coverage of events. 
Existing methods for identifying which events are reported in the news prioritize precision over coverage, since such methods are based on keyword matching \cite{card2015media}, inevitably leading to the lack of generality. 
Third, the measures used to quantify synchrony and diversity of news coverage are not well established. 
To discern the impact of country-level factors on global news coverage, it is crucial to develop robust measures.

We overcome these challenges by developing a novel efficient methodology and applying it to a dataset of 60 million news articles in 10 languages from 124 countries, spanning the first half of 2020. Although the dataset is comprehensive in that it includes global news published in that period, it also includes the onset of the COVID-19 pandemic. This event is uniquely suitable for the study, since it is a rare occurrence of a phenomenon affecting the whole of humanity, and that therefore offers grounds for studying local variations in news coverage.
%
%
%
%
%
%
Our main contributions are: 
\begin{itemize}[topsep=4pt]
    \item a computationally-efficient transformer model that infers multilingual news similarity (\S\ref{sec:transformer_model}),
    \item a similarity-network-based method for identifying global news events from large news corpora (\S\ref{sec:events}),
    \item information-theoretic measures of country-level synchrony and diversity in news coverage of global events (\S\ref{sec:diversity-synchrony-measures}), illustrated in Figure~\ref{fig:intro}, which are superior to baseline measures that do not make use of global news events (\S\ref{sec:diversity-synchrony-baseline}),
    \item by applying the above methods to millions of news articles published across 124 countries, we identify the country-level characteristics that explain international news synchrony and national news diversity globally (\S\ref{sec:multivariate-regression}).
\end{itemize}

Our results have far-reaching implications for understanding systematic differences in media agenda setting and international relations around the globe. 
While limited to the first half of 2020 \textit{only}, we offer unique empirical evidence and explanations for news synchrony and diversity across countries.
Our methodological contributions enable studies of global news coverage across years and millions of news articles, opening the door for a wave of novel agenda-setting studies. 
To bolster the effort of academics and media monitoring agencies, we share the data and code of this study: \url{https://github.com/social-info-lab/global_news_synchrony}.

\section{Related Work}
We contextualize our contributions within a body of related work at the intersection of media studies, computational social science, and machine learning. First, we discuss the societal importance of how news media can set the agenda for the public discourse by deciding what to report on, and how this varies across countries.
Then, we review existing approaches for quantification of media agenda setting, especially looking at the computational methods that afford comparing news at a global scale.

\vspace{-0.25cm}
\subsection{Agenda setting and newsworthiness}
News organizations have to select and prioritize a subset of news to report on each day, a process known as gatekeeping \cite{shoemaker2009gatekeeping}. What the media reports on (and what they do not) has consequences: many studies have shown the ability of the media to ``set the agenda'' \cite{mccombs1972agenda,mccombs2021setting}. That is, the events that media select to report on shapes the public dialogue and helps determine what issues are salient. In this way, the media shape ``not what to think, but what to think about'' \cite{cohen1963press}.
%
%
%
%
Numerous studies have investigated the values by which news organizations decide what stories are ``newsworthy'' enough to report on. Beyond the typical effects of how novel, dramatic, or unusual an event is, the spatial and cultural proximity of the event to a media organization's audience is important \cite{greer2003sex}. 
Beyond studies on which factors shape the choices of individual media organizations, a related body of work investigates how these choices aggregate at a national level to determine patterns of shared media attention and news flow across borders. 
Grasping how news disseminates across different regions is vital for comprehending the global media landscape and its impact on international relations. Mirroring the findings on news newsworthiness, geographic proximity and cultural affinity are found to be important factors, as evidenced by country distance, shared language, religious beliefs, etc. \cite{segev2016group, segev2013america, grasland2020international, wu2003homogeneity, wu2007brave, lee2007international}.
For example, \citet{semmel1977elite} studied the coverage of foreign news by four elite US newspapers, and found that those stories that involved a country that was similar to the US economically, politically, or culturally were more likely to be included.
Additionally, trade has been identified as a potent factor influencing news connections between nations \cite{lee2007international, segev2016group, wu2007brave, guo2020predictors}. Political factors, such as diplomatic relation also play a pivotal role in shaping the news agenda, given that governments can influence journalistic output through regulatory and legislative measures or state-owned media \cite{lee2007international, guo2017global, guo2020predictors}. Due to challenges in analyzing vast amounts of news data, these works remained confined to specific regions. However, computational social science approaches are promising avenues for providing a global perspective about the phenomenon. For example, \citet{zuckerman2003global} leveraged search engine query results for comparing media attention between countries.

\vspace{-0.1cm}
\subsection{News diversity and similarity quantification}


Communication scholarship has a longstanding concern in news story diversity. This body of literature argues that news has become less diverse \cite{noauthor2016-gb,Klinenberg2005-ol}, and pointed to two sources of homogenization: journalistic practice---such as the acceleration of the news cycle and increasing reliance on wire service copy---and the monitoring of other media \cite{bucy2014media, Boczkowski2007-ip}. The rise of Internet and online news reinforces these processes~\cite{mcgregor2019social}, so Internet penetration may contribute to diminished diversity of news coverage~\cite{zuckerman2013rewire}. We test this premise in our analysis of country-level news diversity~(\S\ref{sec:news-diversity}).

Qualitative methods like content analysis, used to derive such findings, enabled gaining deep insights, e.g., about news framing \cite{Shoemaker1996-gp}, while comparing news coverage from multiple outlets \cite[see, e.g.,][]{Beckers2019-py,Choi2009-pc,Scollon2000-bi}. 
However, such qualitative methods are time-consuming and require expertise, thus they are prohibitively costly to perform at a scale, which is necessary for answering questions about news media agenda setting globally. Computational methods are a promising avenue to tackle large-scale analyses \cite{Nicholls2019-gu}. 

Computational research in this field made several attempts at defining similarity between news articles. These definitions opportunistically fitted a computational task, such as automatically detecting whether two articles cover the same real-world happening \cite{Becker2011-og,Gusev2021-tc,Vatolin2021-ja}, story chain \cite{Leskovec2009-ob,Nicholls2019-gu}, event \cite{Xiao2021-zh,Xu2019-ym}, or user preference model \cite{Vrijenhoek2021-zh,Pranjic2020-st,Feng2020-cw}. These tasks enabled theoretical and technological advances, such as deduplicating news stories and recommending personalized news feeds. 
Quantification of news article similarity is a fundamental step that allows comparisons of news article coverage across and within countries. 
Yet, the task-driven definition of similarity of the above body of research limits its applicability to determining the similarity of news articles in a general sense.
Recently, researchers conceptualized and operationalized news similarity as a broader construct that generalizes across event, temporal, and geographic contexts \cite{chen2022semeval}. This broader definition, combined with a large, hand-labeled dataset, enabled NLP researchers to develop machine learning models that achieve human-level performance in estimating multilingual news article similarity.


\begin{figure}[t!]
\centering
\includegraphics[width=0.98\columnwidth]{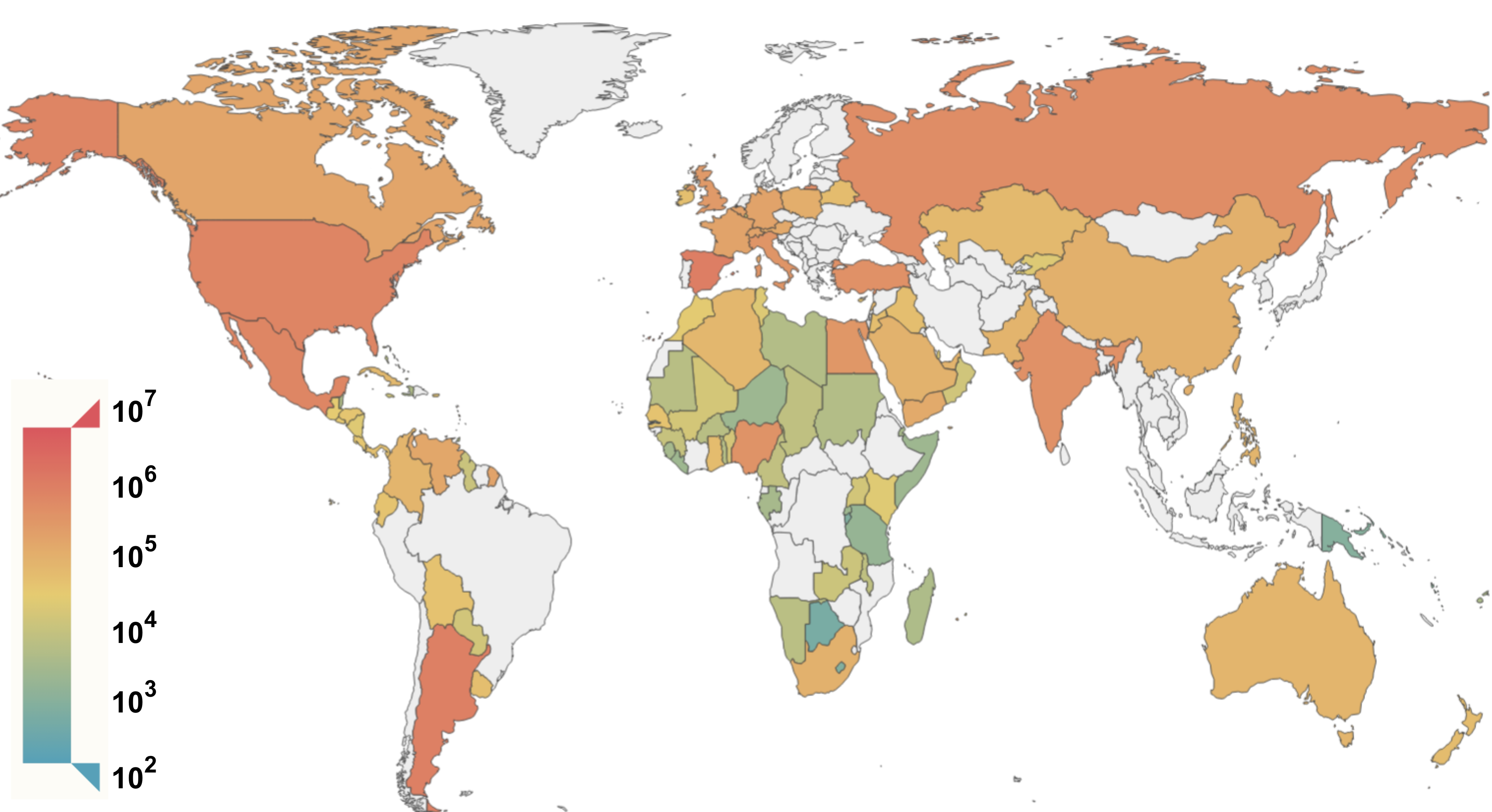} 
\caption{The number of news articles per country. The color bar is scaled logarithmically.}
\label{fig:global_map}
\end{figure}

\subsection{News event detection}
Recent studies also have initiated efforts to detect news events. A common approach involves first obtaining a document representation for each news article \cite{lam2001using, yang2002topic}, e.g. TF-IDF \cite{li2005probabilistic}, and then clustering the news articles based on similarity metrics of their representations \cite{brants2003system, kumaran2004text, huang2016liberal}. Leveraging event detection technologies, extensive event data collections have been established to support societal science research, such as ICEWS \cite{ward2013comparing} and GDELT \cite{leetaru2013gdelt}. These collections offer insights of the details and coverage of news events. 
However, they often limit the event schemes to a \textit{predefined} set, a process that typically demands significant manual effort and is time-consuming in practice. 
Consequently, these schemes may struggle to adapt to new, emerging events over time. Furthermore, these collections are composed of fine-grain events, e.g., shootings. Thus, they do not capture broader journalistic narratives created around such events. 

We take a different approach to news even detection that is not tied to any task or time period, but rather relies on supervised machine learning for news similarity computation and on unsupervised graph clustering for event identification. This approach is computationally efficient and scales well with the number of news articles, enabling a new kind of agenda-setting studies.

\section{Datasets}

To study international news coverage at scale, we collect a large dataset of news articles published in the first half of 2020.
To estimate similarity between news articles from this large dataset, we develop a transformer model by training it on a smaller set of pairs of news articles annotated for their similarity.
%


    

\subsection{Large dataset of news articles}


We use news article data from Media Cloud, which since 2008 queries news feeds of 1.1 million distinct media outlets \cite{roberts2021media}. The platform maintains a list of news outlets organized by country\footnote{\url{https://sources.mediacloud.org/\#/collections/country-and-state}}, which has been carefully curated over years by international media scholars so that it covers important news outlets~\cite{roberts2021media}.

To study the phenomenon world-wide, our dataset includes news in ten languages, chosen for their global significance. We adopt the same ten languages as those in the recent work of \citet{chen2022semeval}, which also allows us to measure news similarity with high accuracy. To ensure the reliability and validity of our news content from a social science perspective, we excluded all URLs from popular social media platforms (such as twitter.com, facebook.com, reddit.com, etc).
In total, we collected metadata and full text of all news articles from January 1, 2020 to June 30, 2020, totaling  $\sim$60M news articles in the following languages: 
English (31M articles),
Spanish (8.2M), 
Russian (7.2M),
German (3.2M),
French (3.2M),
Arabic (2.9M),
Italian (2.4M), 
Turkish (1.1M),
Polish (595K),  
and Mandarin Chinese (342K).

\subsubsection{Annotated news article pairs}


To develop and train a news similarity transformer model, we adopt the dataset and resources provided by a recent study that introduced a nuanced labeling scheme for news similarity~\cite{chen-etal-2022-semeval}. This dataset was extended by additional 21,700 (mostly cross-lingual) news article pairs, which will also be published online.

\subsubsection{Language-agnostic filtering}
We extracted named entities from each news article. As those are expressed in the language of the article, we linked them to respective language-agnostic WikiData entities, by matching them against WikiData titles, echoing techniques found in prior works \cite[e.g.,][]{brank2017annotating}. 
In the reminder of this manuscript, we focus on articles for which we can compute meaningful similarity measures, i.e., those mentioning at least 10 WikiData entities and containing a minimum of 100 words. To avoid the introduction of systematic biases, we estimate English-equivalent length of each article, since the length of the same piece of text differs across languages. This filtering resulted in 15.6 million articles published across 124 countries. This set of countries covers 64\% of all countries and about 76\% the world population (Figure \ref{fig:global_map}).

\begin{figure}[t!]
\centering
\includegraphics[width=0.98\columnwidth]{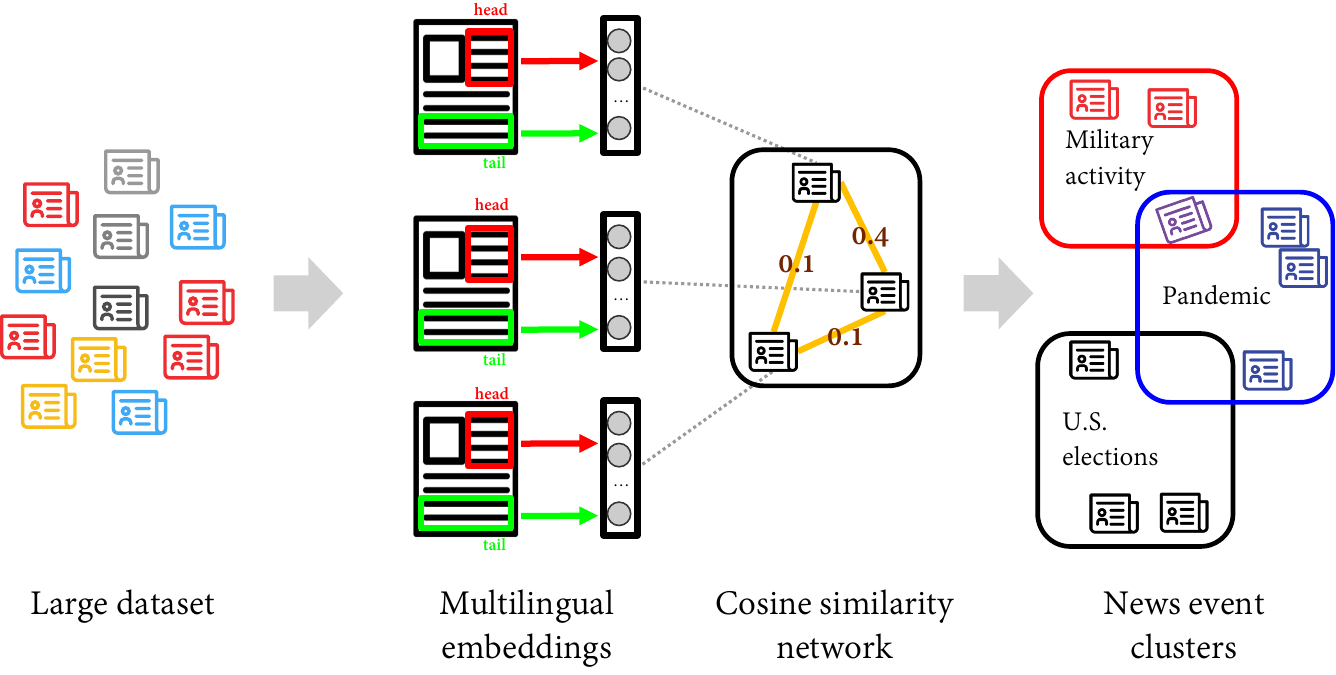} 
\caption{Our pipeline for global news event detection.}
\label{fig:pipeline}
\end{figure}

\section{Global News Event Identification}

We devise the following novel methodology for global news event identification (Figure~\ref{fig:pipeline}). First, we develop an efficient transformer model of pairwise news similarity (\S\ref{sec:transformer_model}). Using this model, we compute similarity between news articles from our large dataset, creating a global news similarity network (\S\ref{sec:news-net}). Then, we cluster this network to identify global news events (\S\ref{sec:events}). Ultimately, we will measure which countries cover each of the events. 




\subsection{Multilingual news article similarity model}
\label{sec:transformer_model}

\setlength{\columnsep}{10pt}
\begin{wraptable}[14]{r}{0.6\columnwidth}
\begin{center}
\vspace{-5pt}
\resizebox{0.6\columnwidth}{!}{
    \begin{tabular}{ccccccccc}
    \toprule
    Architecture & Model & Performance \\
    \toprule
    
    Cross-encoder & {\textbf{HFL}} & 0.811 \\
    \midrule
    \multirow{6}{*}{Bi-encoder} & {MPNet} & 0.492 \\
    
    & {MUSE} & 0.636 \\
    
    & {LaBSE} &  0.680 \\
    
    & {GateNLP} & 0.791\\
    
    
    & {\textbf{This work}} & 0.803 \\
    \bottomrule
    \end{tabular}
    }
    \caption{Performance of pairwise news article similarity models, measured as Pearson correlation on the annotated news dataset, following the prior benchmark~\cite{chen2022semeval}. Best models for each architecture are in bold.}
    \label{tab:model_performance}
\end{center}
\end{wraptable}

Measuring news similarity at a global scale is a challenging task, because of the amount of information included in long-form articles, the inherently cross-lingual setting, and the amount of computation it takes.
To address these challenges, we refined a multilingual deep learning model (LaBSE) to embed the articles and represents them as numerical vectors. The similarity between news articles is then quantified using the cosine similarity between these vectors. We discuss next the model's architecture, and defer implementation details to Appendix A.

Two predominant architectures for text similarity are: (1) cross-encoders which ingest two texts as inputs and directly output a similarity score;  and (2) bi-encoders, which independently transform each input text into a fixed-length vector, and the similarity is then computed as the closeness between these vectors. We favor the bi-encoder approach because of its computational efficiency: representations for bi-encoders can be pre-computed \textit{once} at an $O(n)$ cost and subsequently used for all comparisons. In other words, the computational cost  scales linearly with the number of articles. By contrast, a cross-encoder requires separate computation for each article pair, thus incurring an $O(n^2)$ cost---a computational burden that becomes unsustainable for millions of articles, as in our case. 

Recent research suggested that bi-encoders achieve comparable performance to cross-encoders in assessing news similarity  \cite{chen2022semeval,litterer2023rains}. We confirm this observation by comparing the performance of our model against the state-of-art models \cite{chen2022semeval, singh-etal-2022-gatenlp, conneau2017word, feng2020language} on an updated benchmark developed by \citet{chen2022semeval} (Table \ref{tab:model_performance}).\footnote{The updated benchmark includes the $21,700$ extra news article pairs, i.e., it is more than three times larger than the original benchmark.}
Importantly, our bi-encoder pipeline is an order of magnitude times faster than cross-encoders, which is indispensable for the computation of news similarity across millions of news articles.


\subsection{Model-based similarity computation}
\label{sec:news-net}

Given that our dataset comprises millions of news articles and, hence, many trillions of news article pairs, despite having an efficient transform model for estimating news similarity, it is still computationally infeasible to compute similarity for every potential pair. Fortunately, the vast majority of pairs have nothing in common and can be neglected. 
We take an advantage of this by computing transformer-based similarity only for potentially similar news article pairs.

To identify news article pairs that may be similar, we filter all pairs whose sets of WikiData entities have a Jaccard similarity exceeding 0.25 and were published within a 5-day window of each other. These thresholds are chosen based on the probability that a pair of articles having these characteristics is annotated as similar (Appendix B). This resulted in a final set of 13.6 million news article pairs across 2.2 million unique news articles.
To illustrate the computational challenge of dealing with our dataset, we note that the filtering process \textit{alone} took about a week, utilizing thirty 50-core computing nodes equipped with $200$ GB RAM each. 

While the above filtering step ensures computational viability of the proposed methodology, it focuses on a limited set of 13.6 million news article pairs out of trillions of potential pairs. To address this limitation and to prevent biasing our analysis by the smaller set of pairs, the subsequent steps of our study are based on news event clusters, which encompass a much larger set of 1.4 billion pairs. 
To create a graph of global multilingual news similarity, which will be used to identify news event clusters, we applied the transformer model of news similarity to the filtered news pairs. The news similarity graph represents articles as nodes and their similarity as weighted edges connecting the nodes. 
Most importantly, each of the 1.4 billion pairs contained within the final news event clusters has an impact on our final news diversity and synchrony measures \textit{equivalent} to that of a pair belonging to the 13.6 million set seeding the analysis, as we discuss next.




\begin{figure}[t!]
\centering
\includegraphics[width=0.98\columnwidth]{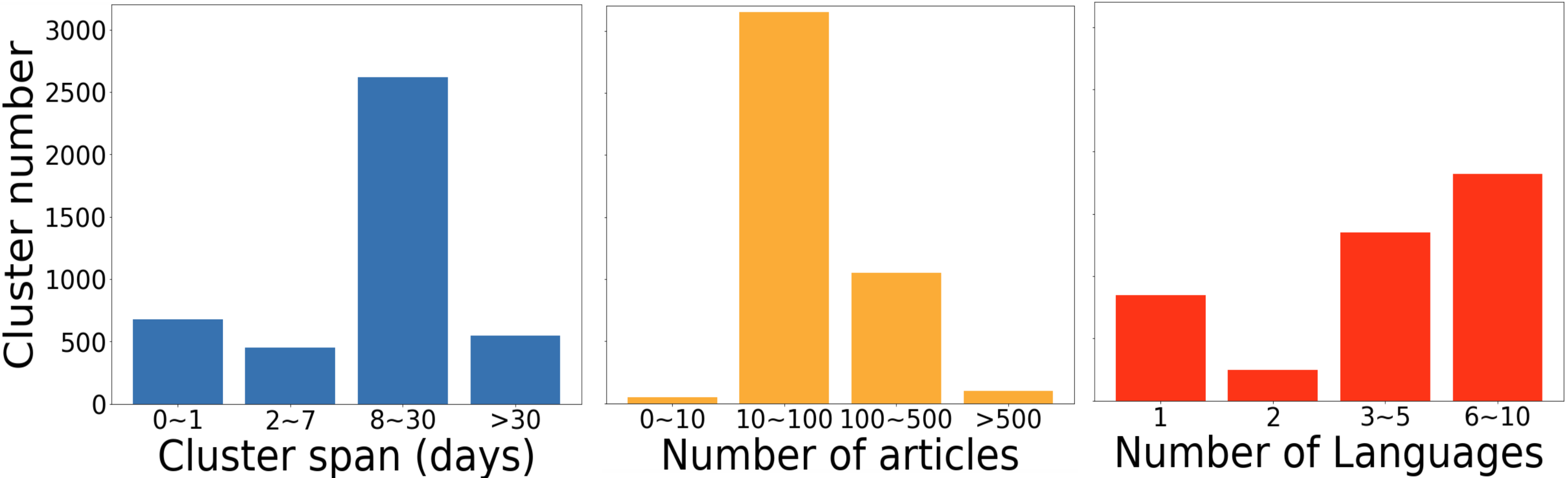}
\caption{The histograms of the characteristics of news events: their duration (left), the number of articles (middle), and the numbers of languages (right).
}
\label{fig:event-stats}
\end{figure}


\begin{figure*}[t!]
\centering
\includegraphics[width=2\columnwidth]{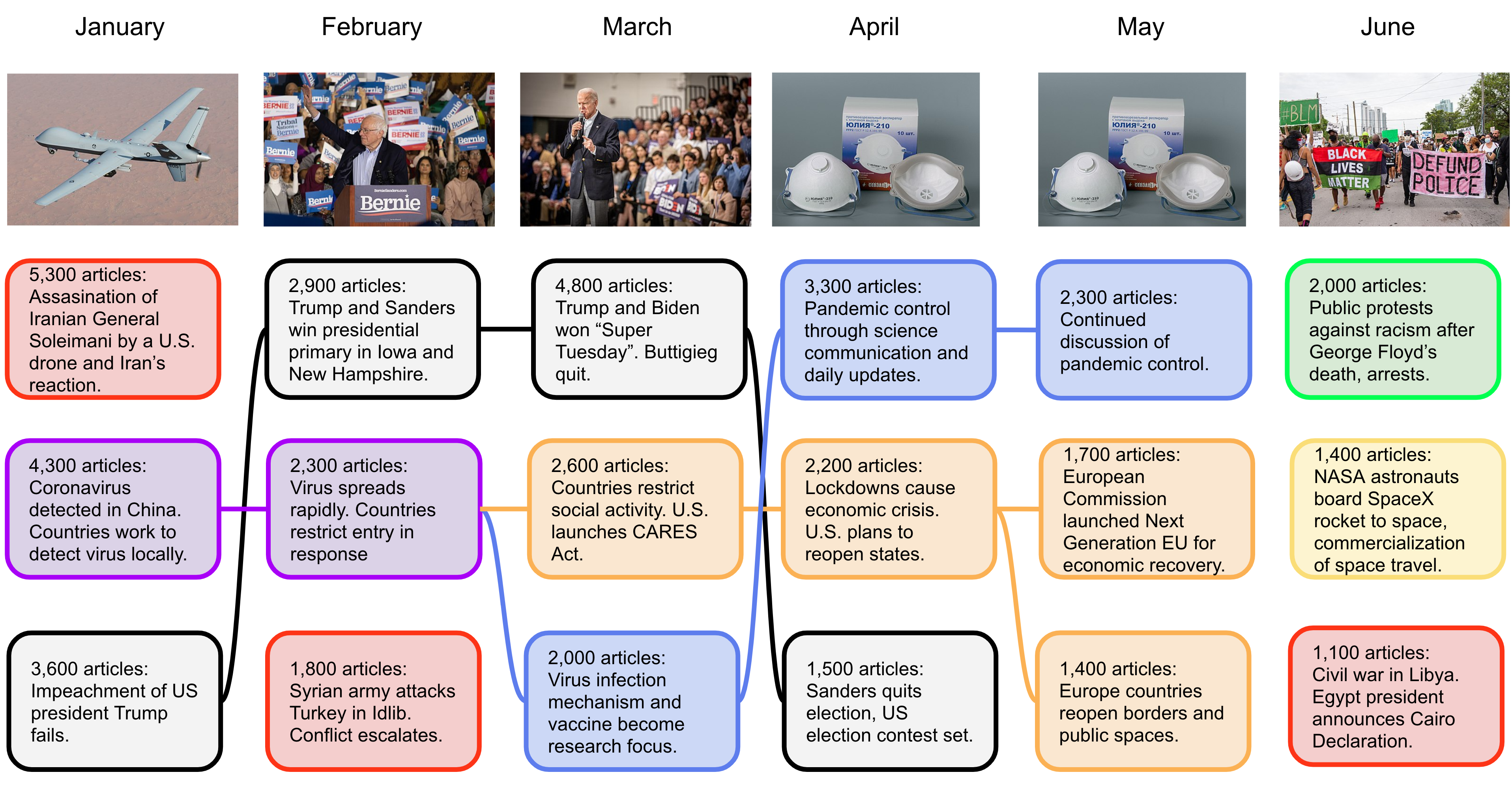} 
\caption{The top 3 largest news event clusters in each month. The majority of these clusters are multilingual and translated into English for interpretation. 
Color corresponds to different branches of an event---U.S Presidential primary elections (black), military operations (red), pandemic (purple) and its two derivative branches: pandemic control (blue) and economic recovery (orange), anti-racism protests (green), and SpaceX rocket launches (yellow). Photos are from Wikipedia. Colors, connections, and descriptions are based on an interpretation of one of the authors. 
}
\label{fig:temporal_event}
\end{figure*}

\subsection{Identification of global news events}
\label{sec:events}

We identify news events by clustering the news similarity network.
Thanks to this step, we move from an \textit{article}-level to an \textit{event}-level analysis of global news.
The event-level description alleviates the limitations of the heuristic filtering we introduced to preserve computational feasibility: article pairs that do not share named entities, for which we did not compute transformer-based similarity, can nevertheless appear in the same news event cluster. Finally, our news diversity and similarity measures treat all news articles within a cluster as equivalent (\S\ref{sec:diversity-synchrony-measures}). In fact, this leads to more accurate model-based explanations of news diversity and synchrony (compare results in \S\ref{sec:diversity-synchrony-baseline} with \S\ref{sec:news-diversity} and \S\ref{sec:news-synchrony}).
Next, we show that news event clusters provide also an interpretable summary of the major happenings during the studied period.


\subsubsection{Clustering method}
To identify news events, we apply the OSLOM algorithm~\cite{lancichinetti2011finding}, which identifies statistically significant, overlapping clusters by optimizing a local fitness function that compares the clusters against random fluctuations in a null network model using order statistics. The method has been successfully used in computational social science~\cite{grabowicz2012social}. 

\subsubsection{Results and evaluation}
\label{sec:event-results}
Overall, we identified 4357 news events. The majority of events lasted between 1 to 4 weeks and were covered by 10 to 100 news articles (Figure~\ref{fig:event-stats}). Around $25\%$ of events were covered by 100 to 500 news articles. Most events spanned multiple languages: nearly half of events were covered in more than 5 languages (out of 10 languages that we studied).




We evaluate the quality of the identified news events. 
To begin with, we find that two articles sharing a cluster are more likely to be similar than two articles sharing a named entity (Appendix C).
To evaluate the coherence of the detected news events, we devised an intrusion task, which is commonly used to evaluate topic models~\cite{chang2009reading}. In this task, human evaluators are presented with 11 articles, all but one sampled from the same cluster. If the ``intruding'' article can be identified consistently, the cluster is considered to be semantically coherent.
Three annotators evaluated 40 clusters. For this evaluation, we chose the 20 largest clusters and 20 random clusters. 
For each news cluster, evaluators were presented with the titles and URLs of the 11 corresponding articles.

We recorded an average precision of $85.8\%$ across the annotators, which is bolstered by substantial inter-rater agreement (0.962 Gwet's AC1 and 0.809 Krippendorff's alpha coefficient). The annotator that spent the most time on the task, achieved $97.5\%$ precision. Compared to analogous results for topic model evaluation~\cite{chang2009reading}, these precision values are relatively high, which lends confidence in the coherence of the news event clusters. 

\subsubsection{News event interpretation}
\label{sec:event-interpretation}

Figure \ref{fig:temporal_event} shows summaries of the top 3 events with the most news articles published each month in the first half of 2020. Consecutive events are sometimes related and can form simple narratives (color-coded in the figure).
Between February and March 2020 the events that attracted the most news articles were the US presidential primaries (black rectangles in the figure). During this period it was unclear who the Democratic party nominee would be. Among Democratic candidates Bernie Sanders led the primaries in February, but Joe Biden later won the key ``Super Tuesday" event in March. Interest in the primaries quickly decreased in April after Sanders left the race leaving Biden as the only competitive candidate for the Democratic nomination.
During that time, news around the COVID-19 pandemic evolved from the identification of the virus in China in January, through realizations of its rapid transmissibility in February and March, to non-pharmaceutical interventions in April and May.

\section{News Coverage Across Countries}\label{sec:multivariate-regression}


Global news event identification enables addressing outstanding questions on how news differ across countries and flow between them. Next, we leverage country characteristics that media scholarship hypothesized as factors determining news coverage (\S\ref{sec:factors}). 
Namely, we use the hypothesized country characteristics as predictors in regression models of \textit{news diversity} within a country and \textit{news synchrony} between countries.
We compare two ways of computing news diversity and synchrony: a naive baseline approach that only uses global news similarity network (\S\ref{sec:diversity-synchrony-baseline}) and our main approach that uses global news events and treats articles within a news event cluster as discussing the same happenings (\S\ref{sec:diversity-synchrony-measures}). In comparison to the former, the latter method results in superior explanatory power of models regressing news synchrony and diversity against country-level characteristics.

\subsection{Regression structure and representativeness}
For each country and country pair we measure news diversity and synchrony, respectively. Then, we regress these measures against predictors described next.
To mitigate potential news data biases across countries, we weight all countries equally. For our news diversity models, we ensure an equal number of news articles is sub-sampled from each of the 124 countries. Meanwhile, in our news synchrony models, we use 7626 country pairs as our samples.

\subsection{Feature selection and preprocessing}
\label{sec:factors}

\begin{wraptable}[25]{r}{0.5\columnwidth}
\vspace{-0.3cm}
\resizebox{0.5\columnwidth}{!}{
    \NoHyper
    \begin{tabular}{r p{0.6\columnwidth}}
    \\ 
    \toprule
    Type & Predictor \\
    \toprule
   \multirow{3}{*}{Econ.} & \text{{\color{orange} Trade}}  \cite{lee2007international, kariel1984factors, segev2016group, wu2007brave, dupree1971international, guo2020predictors} \\    
            & \text{{\color{orange} GDP}}  \cite{lee2007international, segev2015visible, segev2016group, grasland2020international, guo2020predictors} \\
            & \text{{\color{orange} Financial Investment}}  \cite{lee2007international} \\
    \hline
    \multirow{7}{*}{Pol.} & \text{{\color{blue} Democracy index}}  \cite{ robinson1976international, nnaemeka1981internal, kim1996determinants, segev2016group} \\
                         & \text{{\color{blue} Political blocs}}  \\
                         & \text{{\color{blue} Press freedom}}  \cite{wu1998systemic, wu2007brave} \\ 
                         &  \text{{\color{blue} Mode of government}}  \cite{hester1973theoretical, wilkinson2012trending} \\ 
                         & \text{{\color{blue} Diplomatic ties}}  \cite{lee2007international, guo2017global, guo2020predictors} \\
                         & \text{{\color{blue} Military strength index}}  \cite{wu2007brave, segev2015visible, kim1996determinants, guo2020predictors} \\
                         & \text{{\color{blue} Conflict intensity (peace index)}}  \cite{segev2015visible, segev2016group} \\
    \hline
    \multirow{4}{*}{Soc.} &  \text{{\color{brown} Population}}   \cite{lee2007international, kariel1984factors, segev2016group, grasland2020international, kim1996determinants, dupree1971international} \\
                        & \text{{\color{brown} Immigration}}   \cite{wu2007brave, segev2015visible, segev2016group, jansen2019drives} \\
                        & \text{{\color{brown} Population density}}   \cite{dupree1971international} \\ 
                        & \text{{\color{brown} Gini index}}   \cite{segev2016group} \\
    \hline
    \multirow{4}{*}{Geo.} & \text{{\color{green} Distance}}  \cite{kariel1984factors, segev2016group, segev2013america, grasland2020international, kim1996determinants, dupree1971international} \\
                          & \text{{\color{green} Neighbors}}   \cite{grasland2020international, kariel1984factors, segev2015visible, wu2007brave, guo2020predictors} \\
                          & \text{{\color{green} Area}}  \cite{lee2007international, segev2016group, grasland2020international} \\
                          & \text{{\color{green} Continent}}  \cite{kim1996determinants, dupree1971international, guo2020predictors}   \\
    \hline
    \multirow{5}{*}{Cul.} & \text{{\color{purple} Language}}  \cite{kariel1984factors, segev2016group, segev2013america, wu2007brave, kim1996determinants, dupree1971international} \\
                        & \text{{\color{purple} Religion}}  \cite{lee2007international, segev2016group, segev2013america} \\
                        & \text{{\color{purple} Media outlets number}}  \cite{wu2003homogeneity, wu2007brave, segev2016group, gerbner1977many}  \\
                        & \text{{\color{purple} Literacy rate}}  \cite{lee2007international, segev2016group, dupree1971international} \\ 
                        & \text{{\color{purple} Internet user rate}}  \\
    \bottomrule
    \end{tabular}
    }
    \caption{Predictors considered in the regressions of news diversity and synchrony, including the references to studies that mention them. We have not identified references mentioning political blocs and Internet user rate.}
    \label{tab:country_factors}
\end{wraptable}

We surveyed extensive literature for traits of countries and international relations that are related to news practices, and categorized them into five types: economic, political, societal, geographic and cultural \cite{lee2007international, segev2016group}. All considered factors are listed (color-coded by the type) in Table \ref{tab:country_factors} with detailed explanation and literature review in Appendix D.  In addition to factors identified by the literature, we took into account two additional factors: the Internet user rate, to account for the relevance of online news in today's media landscape, and whether countries are members of three major blocs---NATO, EU, and BRICS---given their importance in geopolitics. Country indices are sourced from the World Bank and The Organisation for Economic Co-operation and Development (OECD).

\subsubsection{Feature preprocessing}
Each factor is either numeric or categorical. Categorical factors are represented as dummy vectors. For news synchrony analysis, we categorize all numeric factors. Subsequently, the category combination of a country pair is treated as the pairwise category and is also represented using dummy vectors.
We re-scaled all numeric predictors and the dependent variables (diversity and synchrony) so that they take values between 0 and 1. 

\subsubsection{Feature and model selection}
To avoid co-linearity, we applied feature and model selection techniques. Namely, we employ two well-known criteria---Variance Inflation Factor (VIF) and Akaike information criterion (AIC)---and report results for the best models according to each criterion. We report models constructed with either of the techniques to showcase robustness of our results and methodology. VIF is a popular feature selection technique, whereas AIC is a model selection method that identifies the predictors that best explain the dependent variables.
For all regressions we apply the same model and feature selection.

\subsection{Baseline synchrony and diversity regression}
\label{sec:diversity-synchrony-baseline}

A naive measure of news article diversity or similarity could be the average similarity, as computed by the transformer model, of articles that were published within a country or between two countries, respectively.
These measures, however, face a few issues. 
First, it is computationally infeasible to compute similarity between all news article pairs. We were able to compute news article similarity for only 13.6 million news article pairs, which constitute only 1\% of all pairs contained in global event clusters. 
Second, it is not clear whether such simple averages are good measures of news diversity and synchrony.

To showcase these issues and their consequences, we first perform our analysis using this baseline approach, without relying on the global news event clusters. 
Following this approach, to explain diversity of news, we regress the average pairwise similarity of news published in that country against the country characteristics. To explain news synchrony, we regress the average similarity of news published between a country pair. 
The resulting regression models, selected by optimizing the AIC,  achieved low adjusted $R^2=0.13$ and $R^2=0.30$ for news diversity and synchrony, respectively. By contrast, our main method based on global events (described in the next subsection) achieved adjusted $R^2=0.536$ and $R^2=0.448$, respectively, which is notable and significantly surpasses that of similar works in social science studies \cite{wu2007brave, guo2020predictors}.
In other words, the baseline models explain much less variance in the similarity and diversity measures than the models based on global events. That said, when comparing the coefficients of the sets of models, we find qualitatively similar results (Appendix E). 



\subsection{News diversity and synchrony measures}
\label{sec:diversity-synchrony-measures}

For each global news event, we compute the percentage of a country's news articles reporting on it. In this way, we form a distribution of articles over news events for each country. Then, we define the \textit{news diversity} within a country as Shannon's entropy of this country's distribution \cite{shannon1948mathematical}. Higher entropy indicates greater diversity.
To measure the similarity of the events covered by news media across countries, we introduce \textit{news synchrony}.
We define news synchrony between a pair of countries as the negative Jensen-Shannon divergence of their respective event distributions \cite{fuglede2004jensen}. The more negative the divergence, the more synchronized the countries' media. We chose the term news synchrony over similarity because each cluster encompasses a specific temporal span, and thus captures both semantic and temporal alignment of news coverage.

By anchoring the news diversity and synchrony measures in global news events, we progress from an analysis rooted in 13.6 million news article pairs to an analysis encompassing 1.4 billion article pairs that are within clusters. Namely, even if a pair of articles did not pass the initial filtering (\S\ref{sec:news-net}), they can still appear in the same news event cluster (there are 1.4 billion of such pairs) and have the same impact on news diversity and synchrony measures as pairs that did pass the filtering and ended up within a cluster (there are only 13.6 million of these pairs). The reason for this is that our news diversity and synchrony measures depend only on the distribution of articles across news event clusters, i.e., two articles that appear in the same cluster are treated as equivalent.

%


    

\begin{figure}[t!]
\centering
\includegraphics[width=\columnwidth]{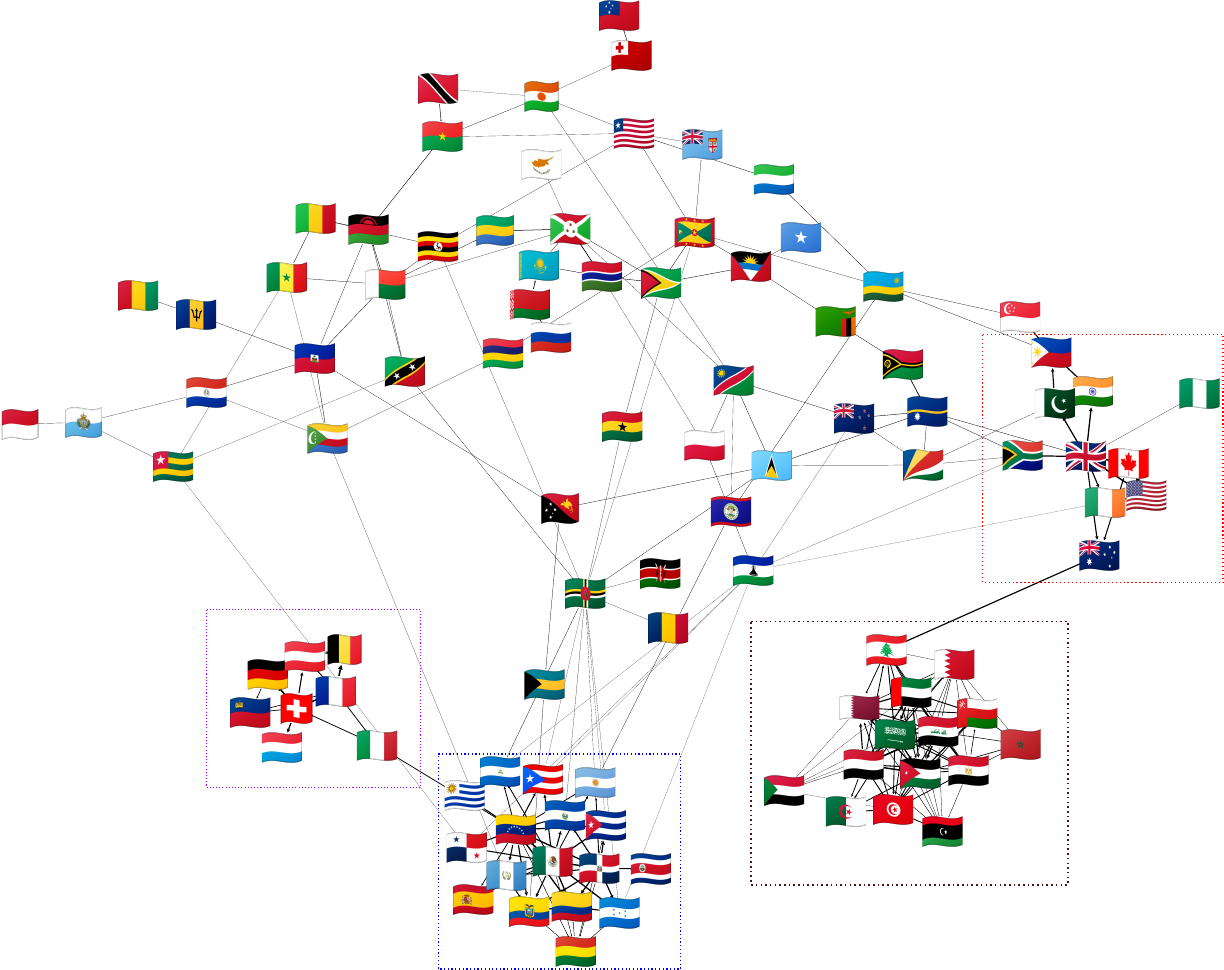}
\caption{News synchrony graph backbone for the top 100 countries with the largest populations. Each edge represents 95\% confidence that the respective news synchrony is non-random \cite{serrano2009extracting}. Some countries are not selected into the graph backbone, e.g. China. The main clusters are marked with squares: the US and the UK and its former colonies (red square), 5 countries of the old European Union of 1958 (purple), Latin America and Spain (blue), the Arab world (brown).}
\label{country_graph}
\end{figure}


\begin{table}[tb!]
\begin{center}
\resizebox{0.4\textwidth}{!}{ 
    \begin{tabular}{ccr@{}lr@{}l}
    \toprule
    Type & Predictor & \makecell{VIF model\\coefficient} & &  \makecell{AIC model\\coefficient} & \\
    \toprule
    
    & {\color{gray} Intercept} & $0.347$ && $0.414$  \\
    \midrule
    
    Pol. & {\color{blue} Federalism?} & $0.118$ &$^{**}$  & $0.069$ &$^{**}$ \\
    
    & {\color{blue} Within NATO?} & $0.073$ &$^{**}$ & $0.044$ &$^{**}$ \\
    & {\color{blue} Within EU?} & $-0.028$ &$^{**}$ & $-0.093$ &$^{*}$   \\
    & {\color{blue} Within BRICS?} & $0.006$ &$^{**}$  & $0.012$ &$^{**}$   \\
    
    & {\color{blue} Democracy index} &  && $0.175$ &$^{**}$  \\
    & {\color{blue} Peace index} &  && $-0.195$ &$^{**}$  \\
    
    \midrule
    Soc. & {\color{brown} Gini index} & $0.174$ &$^{**}$ & $0.196$ &$^{**}$ \\
    
    & {\color{brown} Population} & $0.146$ &$^{*}$ & $0.102$ &$^{*}$ \\
    & {\color{brown} Net Migration} &&  & $0.027$ &$^{**}$\\
    
    \midrule
    Geo. & {\color{green} Area} && & $0.093$ &$^{**}$\\
    \midrule
    
    Cul. & {\color{purple} \#media} & $-0.148$ &$^{**}$ & $-0.130$ &$^{**}$    \\
    & {\color{purple} Same media?} & $-0.069$ &$^{**}$ & $-0.063$ &$^{**}$  \\
    
    & \textbf{{\color{purple} \#languages}} & $\textbf{0.212}$ &$^{**}$ & $\textbf{0.250}$ &$^{**}$   \\
    & {\color{purple} \#language families} & $0.058$ &$^{**}$ & $0.038$ &$^{**}$  \\
    & {\color{purple} Religion diversity} & $0.154$ &$^{**}$ & $0.126$ &$^{**}$ \\
    & {\color{purple} \textbf{Internet user ratio}} & $\textbf{0.300}$ &$^{**}$ & $\textbf{0.371}$ &$^{**}$ \\ 
    & {\color{purple} Literacy Ratio} &&  & $-0.126$ &$^{**}$ \\
    \bottomrule
    \end{tabular}
}
\caption{News diversity regression coefficients. Positive coefficients mean more news diversity. Predictor color corresponds to its category. The top 2 predictors are in bold. Missing predictors or values indicate they are dropped during feature selection process. ** and * symbols stand for 0.005 and 0.01 confidence intervals respectively. }
\label{tab:intra_country_model_result}
\end{center}
\end{table}

\noindent

\subsection{News diversity within countries}
\label{sec:news-diversity}

The five countries with the highest internal news diversity are the U.S.A., the United Kingdom, Canada, Switzerland, and Mexico.
To explain the news diversity of each country, we regress it against country-level characteristics.
The adjusted R$^2$ of the regression models are 0.494 (for the best model with factors selected through VIF) and 0.536 (for the best model selected via AIC). Table \ref{tab:intra_country_model_result} lists the coefficients (all statistically significant, $p<0.01$) of the factors. 

The results indicate that the Internet user rate within a country is the strongest predictor of its news diversity ($p<0.005$). 
The higher the Internet penetration, the larger the news diversity. This result is remarkable, given that prior studies suggest that news diversity diminishes due to the acceleration of the news cycle, facilitated by the rise of 24-hour news cycle and online news~\cite{bucy2014media, Boczkowski2007-ip, mcgregor2019social, zuckerman2013rewire}. In contrast, we find that Internet adoption diversifies news coverage within a population, probably because news media cater to diverse interests and motivations of online audiences \cite{lee2013news}.

The second strongest predictor is the number of official languages. Intuitively, the more languages and language families in a country, the higher news diversity ($p<0.005$). Similarly, religious diversity, measured as the Shannon entropy of the distribution over religions in a country, contributes to news diversity. We conclude, in agreement with prior works~\cite{kariel1984factors, segev2016group, segev2013america, wu2007brave, kim1996determinants, dupree1971international, lee2007international}, that cultural diversity and news diversity are closely related.

Among societal factors, a larger population ($p<0.01$) and a higher Gini index ($p<0.005$) are related to more diverse interests and news. 
Gini index indicates income inequality of the population and different incomes correspond to different news consumption patterns \cite{hughes2011media, alexander2018digital}.
Among political factors, federalist government is a predictor of diverse news, because the policy aims of state governments are often different from the national government \cite{baumgartner2010agendas}, decentralizing media focus. 

\begin{table}[tb!]
\resizebox{0.98\columnwidth}{!}{
    \centering

    \begin{tabular}{ccr@{}lr@{}l}
    \toprule
    Type & Predictor & \makecell{VIF model\\coefficient} &&  \makecell{AIC model\\coefficient} & \\
    \midrule
    & {\color{gray} Intercept} & $0.238$ && $0.247$ \\
    \midrule
    Econ. & {\color{orange} \textbf{Trade}} & $\textbf{0.132}$&$^{*}$& $\textbf{0.155}$&$^{**}$\\
    & {\color{orange} GDP (high--high)} & $0.073$&$^{*}$& $0.073$&$^{*}$ \\
    & {\color{orange} GDP (low--high)} & $0.011$&$^{**}$ &   \\
    \midrule
    Pol. & {\color{blue} Same government mode} & $0.010$&$^{*}$& \\
    & {\color{blue} Both federalism} & $0.059$&$^{*}$& $0.057$&$^{**}$\\
    & {\color{blue} Federalism and other} & $0.010$&$^{*}$&  \\

    & {\color{blue} Both in NATO} & $0.051$&$^{*}$& $0.035$&$^{*}$\\
    & {\color{blue} Both in BRICS} & $0.095$&$^{*}$& $0.086$&$^{*}$\\
    
    & {\color{blue} Across NATO--BRICS groups} & $0.046$&$^{*}$& $0.060$&$^{*}$\\
    \midrule

    Soc. & {\color{brown} Same Gini index category} & $0.020$&$^{*}$& \\
    & {\color{brown} Gini index (high--high)} & $-0.037$&$^{*}$& $-0.042$&$^{**}$\\
    & {\color{brown} Gini index (low--high)} & $-0.018$&$^{*}$& \\

    & {\color{brown} Democracy index (high--high)} & $0.076$&$^{*}$& $0.068$&$^{**}$\\
    & {\color{brown} Democracy index (low--high)} & $0.023$&$^{*}$& \\
    \midrule
    
    Geo. & {\color{green} Neighbor} & $0.016$&$^{*}$&\\
    & {\color{green} Same Continent} & $0.012$&$^{*}$& $0.022$&$^{**}$\\
    \midrule
    
    Cul. & \textbf{{\color{purple} Same language}} & $\textbf{0.125}$&$^{*}$& $\textbf{0.139}$&$^{**}$\\
    & {\color{purple} Speak English} & $-0.118$&$^{*}$& $-0.130$&$^{*}$\\
    & {\color{purple} Speak German} & $0.014$&$^{*}$&  \\
    & {\color{purple} Speak French} & $-0.079$&$^{*}$& $-0.084$&$^{*}$\\
    & {\color{purple} Speak Arabic} & $0.026$&$^{*}$&  \\
    & {\color{purple} Speak Italian} & $-0.135$&$^{*}$& $-0.125$&$^{*}$\\
    & {\color{purple} Speak Russian} & $-0.036$&$^{*}$&  \\

    \bottomrule
    \end{tabular}
    }
    
    \caption{News synchrony regression coefficients. Positive coefficients mean more news synchrony between a pair of countries. 
    GDP is binarized as ``high" (\textgreater \$500 billion) and ``low" (\textless \$500 billion). Democracy index  is encoded as ``high" (\textgreater 5) and ``low" (\textless 5).
    Colors and stars follow Table~\ref{tab:intra_country_model_result}.
    }
    \label{tab:inter_country_different_metrics}
\end{table}

\subsection{News synchrony between countries}
\label{sec:news-synchrony}

Figure \ref{country_graph} depicts the backbone of the international news synchrony network, extracted using an algorithm identifying non-random edges~\cite{serrano2009extracting}. In our case, edges indicate synchrony between the countries. 
The graph reveals four recognizable groups: Latin America, the Arab world, five countries of the old European Union of 1958 (plus Lichtenstein, Austria, and Switzerland), and a group including the US, the UK, and its former colonies. 
These groups align with the geographic-linguistic communities in \cite{kim1996determinants}, which examined the structure of international news flow by regressing the number of international newspapers and trade amount between countries against country characteristics.
The most-connected countries include France, Canada, Belgium, Mexico, Switzerland, and the United Kingdom. These nations consistently rank among the top 10 when evaluated using both PageRank \cite{page1999pagerank} and betweenness centrality. 

Next, we regress news synchrony against the characteristics of each country pair (Table \ref{tab:inter_country_different_metrics}). The adjusted R$^2$ of the models are 0.446 (VIF) and 0.448 (AIC). Trade volume is the strongest predictor, which corroborates prior findings \cite{wu2000systemic,segev2016group}. 
While sharing a common language is associated with higher news synchrony in general, this effect varies for specific languages.\footnote{The effect for a particular country can be obtained by summing the "Same language" coefficient with the coefficient of the particular language.}
On the one hand, countries speaking German or Arabic show more news synchrony, which might be attributed to their similarities in values, lifestyles, and social norms.
On the other hand, there is much less news synchrony between countries speaking English or French, likely due to the diversity of countries officially using these languages. Speaking Italian and Russian also correlates with less news synchronization. Italian-speaking countries are relatively idiosyncratic, e.g., Switzerland is arguably not influenced much by Italy, while Vatican has a very particular role. On the other hand, Russian-speaking countries, by virtue of recent conflicts, may have different news in their focus. 

Sharing a border, or being broadly located in the same continent, intuitively correlates with news synchrony. However, controlling for these factors, geographic distance does not show a significant effect on news synchrony---a result that may be attributed to transportation, globalization, and the role of the Internet \cite{guo2020predictors}. The countries with the same government mode are more likely to synchronize with each other in the news event coverage, especially those with federalist systems, which may indicate that these countries share public interests, ideology, or face similar social issues. 

Interestingly, countries that belong to NATO experience more news synchrony, possibly because they have common security concerns and some of the largest events correspond to military operations (Figure~\ref{fig:temporal_event}). Countries belonging to BRICS exhibit more synchronized news, possibly due to their common developmental interests. A similar effect is not evident for the EU members, which may be the result of the recent resurgence of nationalist feelings among its member states.

\section{Conclusions}


This work provides empirical evidence that substantiates our understanding of the global news ecosystem, and unpacks the factors that influence news coverage within and across countries. 
%
On the one hand, we find that population size, inequality, linguistic and religious diversity tend to be related with more diverse news coverage. On the other hand, nations that are proximate geographically, culturally, and socially, maintaining economic and political relations, tend to synchronize their coverage of news events, which likely further reinforces their connection. 

The proposed methodology is a significant step towards systematic study and discovery of unexpected phenomena in global news coverage. 
For instance, prior studies suggest that the acceleration of the news cycle in the Internet age contributes to the homogenization of news coverage~\cite{bucy2014media, Boczkowski2007-ip, mcgregor2019social, zuckerman2013rewire}. However, here we found that Internet adoption is the \textit{strongest} predictor of news diversity within a country ($p<0.005$). This finding illustrates the potential of the proposed methodology to contribute to communication science in ways that go beyond confirmations of existing theories.


%
Our methodological contributions enable studies of global news coverage across years and millions of news articles, opening the floodgates for a wave of novel agenda-setting studies. 
For instance, media scholars signaled a decline in global news diversity \cite{noauthor2016-gb,Klinenberg2005-ol, zuckerman2013rewire}. This hypothesis can be tested by measuring news diversity over tens of years.
Future works can identify and explain the dynamics of biases in global news coverage (e.g., 40 times as many people must die in a disaster in Africa to get the same level of U.S. news coverage as a disaster in Eastern Europe~\cite{eisensee2007news}).
Particularly interesting is the analysis of the long-term dynamics in news coverage across countries in the periods preceding war outbreaks. Such research could offer a more comprehensive understanding of the complex interplay between media, public opinion, politicians, international relations, and global conflict outbreaks. 
Taking a broader perspective, international news coverage reflects complex geopolitical histories and local realities. As such, this work offers methodologies to observe and interpret our continuously evolving global narrative, as seen through the lens of the news that document it.

\section*{Acknowledgments}
This research has received funding through grants from the Volkswagen Foundation and the University of Massachusetts Amherst. We sincerely thank Media Cloud for data access and the annotators for annotating multilingual news similarity.

\bibliographystyle{ACM-Reference-Format}
\bibliography{refs,aaai22}


\begin{thebibliography}{93}


\ifx \showCODEN    \undefined \def \showCODEN     #1{\unskip}     \fi
\ifx \showDOI      \undefined \def \showDOI       #1{#1}\fi
\ifx \showISBNx    \undefined \def \showISBNx     #1{\unskip}     \fi
\ifx \showISBNxiii \undefined \def \showISBNxiii  #1{\unskip}     \fi
\ifx \showISSN     \undefined \def \showISSN      #1{\unskip}     \fi
\ifx \showLCCN     \undefined \def \showLCCN      #1{\unskip}     \fi
\ifx \shownote     \undefined \def \shownote      #1{#1}          \fi
\ifx \showarticletitle \undefined \def \showarticletitle #1{#1}   \fi
\ifx \showURL      \undefined \def \showURL       {\relax}        \fi
\providecommand\bibfield[2]{#2}
\providecommand\bibinfo[2]{#2}
\providecommand\natexlab[1]{#1}
\providecommand\showeprint[2][]{arXiv:#2}

\bibitem[Alexander et~al\mbox{.}(2018)]%
        {alexander2018digital}
\bibfield{author}{\bibinfo{person}{Victoria~D Alexander},
  \bibinfo{person}{Grant Blank}, {and} \bibinfo{person}{Scott~A Hale}.}
  \bibinfo{year}{2018}\natexlab{}.
\newblock \showarticletitle{Digital traces of distinction? Popular orientation
  and user-engagement with status hierarchies in TripAdvisor reviews of
  cultural organizations}.
\newblock \bibinfo{journal}{\emph{New Media \& Society}} \bibinfo{volume}{20},
  \bibinfo{number}{11} (\bibinfo{year}{2018}), \bibinfo{pages}{4218--4236}.
\newblock
\urldef\tempurl%
\url{https://doi.org/10.1177/1461444818769448}
\showDOI{\tempurl}


\bibitem[Baum and Zhukov(2015)]%
        {baum2015filtering}
\bibfield{author}{\bibinfo{person}{Matthew~A Baum} {and}
  \bibinfo{person}{Yuri~M Zhukov}.} \bibinfo{year}{2015}\natexlab{}.
\newblock \showarticletitle{Filtering revolution: Reporting bias in
  international newspaper coverage of the Libyan civil war}.
\newblock \bibinfo{journal}{\emph{Journal of Peace Research}}
  \bibinfo{volume}{52}, \bibinfo{number}{3} (\bibinfo{year}{2015}),
  \bibinfo{pages}{384--400}.
\newblock
\urldef\tempurl%
\url{https://doi.org/10.1177/0022343314554791}
\showDOI{\tempurl}
\showeprint{https://doi.org/10.1177/0022343314554791}


\bibitem[Baumgartner and Jones(2010)]%
        {baumgartner2010agendas}
\bibfield{author}{\bibinfo{person}{Frank~R Baumgartner} {and}
  \bibinfo{person}{Bryan~D Jones}.} \bibinfo{year}{2010}\natexlab{}.
\newblock \bibinfo{booktitle}{\emph{Agendas and instability in American
  politics}}.
\newblock \bibinfo{publisher}{University of Chicago Press}.
\newblock


\bibitem[Becker et~al\mbox{.}(2011)]%
        {Becker2011-og}
\bibfield{author}{\bibinfo{person}{Hila Becker}, \bibinfo{person}{Mor Naaman},
  {and} \bibinfo{person}{Luis Gravano}.} \bibinfo{year}{2011}\natexlab{}.
\newblock \showarticletitle{Beyond Trending Topics: {Real-World} Event
  Identification on Twitter}.
\newblock \bibinfo{journal}{\emph{ICWSM}} \bibinfo{volume}{5},
  \bibinfo{number}{1} (\bibinfo{year}{2011}), \bibinfo{pages}{438--441}.
\newblock


\bibitem[Beckers et~al\mbox{.}(2019)]%
        {Beckers2019-py}
\bibfield{author}{\bibinfo{person}{Kathleen Beckers}, \bibinfo{person}{Andrea
  Masini}, \bibinfo{person}{Julie Sevenans}, \bibinfo{person}{Miriam van~der
  Burg}, \bibinfo{person}{Julie De~Smedt}, \bibinfo{person}{Hilde Van~den
  Bulck}, {and} \bibinfo{person}{Stefaan Walgrave}.}
  \bibinfo{year}{2019}\natexlab{}.
\newblock \showarticletitle{Are newspapers' news stories becoming more alike?
  Media content diversity in Belgium, 1983--2013}.
\newblock \bibinfo{journal}{\emph{Journalism}} \bibinfo{volume}{20},
  \bibinfo{number}{12} (\bibinfo{date}{Dec.} \bibinfo{year}{2019}),
  \bibinfo{pages}{1665--1683}.
\newblock


\bibitem[Boczkowski and de~Santos(2007)]%
        {Boczkowski2007-ip}
\bibfield{author}{\bibinfo{person}{Pablo~J Boczkowski} {and}
  \bibinfo{person}{Martin de Santos}.} \bibinfo{year}{2007}\natexlab{}.
\newblock \showarticletitle{When More Media Equals Less News: Patterns of
  Content Homogenization in Argentina's Leading Print and Online Newspapers}.
\newblock \bibinfo{journal}{\emph{Political Communication}}
  \bibinfo{volume}{24}, \bibinfo{number}{2} (\bibinfo{date}{May}
  \bibinfo{year}{2007}), \bibinfo{pages}{167--180}.
\newblock


\bibitem[Boomgaarden et~al\mbox{.}(2010)]%
        {boomgaarden2010news}
\bibfield{author}{\bibinfo{person}{Hajo~G Boomgaarden}, \bibinfo{person}{Rens
  Vliegenthart}, \bibinfo{person}{Claes~H De~Vreese}, {and}
  \bibinfo{person}{Andreas~RT Schuck}.} \bibinfo{year}{2010}\natexlab{}.
\newblock \showarticletitle{News on the move: Exogenous events and news
  coverage of the European Union}.
\newblock \bibinfo{journal}{\emph{Journal of European Public Policy}}
  \bibinfo{volume}{17}, \bibinfo{number}{4} (\bibinfo{year}{2010}),
  \bibinfo{pages}{506--526}.
\newblock


\bibitem[Bourdieu(1993)]%
        {bourdieu1993field}
\bibfield{author}{\bibinfo{person}{Pierre Bourdieu}.}
  \bibinfo{year}{1993}\natexlab{}.
\newblock \bibinfo{booktitle}{\emph{The Field of Cultural Production}}.
\newblock \bibinfo{publisher}{Columbia University Press}.
\newblock


\bibitem[Brank et~al\mbox{.}(2017)]%
        {brank2017annotating}
\bibfield{author}{\bibinfo{person}{Janez Brank}, \bibinfo{person}{Gregor
  Leban}, {and} \bibinfo{person}{Marko Grobelnik}.}
  \bibinfo{year}{2017}\natexlab{}.
\newblock \showarticletitle{Annotating documents with relevant wikipedia
  concepts}.
\newblock \bibinfo{journal}{\emph{Proceedings of SiKDD}}  \bibinfo{volume}{472}
  (\bibinfo{year}{2017}).
\newblock


\bibitem[Brants et~al\mbox{.}(2003)]%
        {brants2003system}
\bibfield{author}{\bibinfo{person}{Thorsten Brants}, \bibinfo{person}{Francine
  Chen}, {and} \bibinfo{person}{Ayman Farahat}.}
  \bibinfo{year}{2003}\natexlab{}.
\newblock \showarticletitle{A system for new event detection}. In
  \bibinfo{booktitle}{\emph{Proceedings of the 26th annual international ACM
  SIGIR conference on Research and development in informaion retrieval}}.
  \bibinfo{pages}{330--337}.
\newblock


\bibitem[Bucy et~al\mbox{.}(2014)]%
        {bucy2014media}
\bibfield{author}{\bibinfo{person}{Erik~P Bucy}, \bibinfo{person}{Walter
  Gantz}, {and} \bibinfo{person}{Zheng Wang}.} \bibinfo{year}{2014}\natexlab{}.
\newblock \showarticletitle{Media technology and the 24-hour news cycle}.
\newblock In \bibinfo{booktitle}{\emph{Communication technology and social
  change}}. \bibinfo{publisher}{Routledge}, \bibinfo{pages}{143--163}.
\newblock


\bibitem[Card et~al\mbox{.}(2015)]%
        {card2015media}
\bibfield{author}{\bibinfo{person}{Dallas Card}, \bibinfo{person}{Amber
  Boydstun}, \bibinfo{person}{Justin~H Gross}, \bibinfo{person}{Philip Resnik},
  {and} \bibinfo{person}{Noah~A Smith}.} \bibinfo{year}{2015}\natexlab{}.
\newblock \showarticletitle{The media frames corpus: Annotations of frames
  across issues}. In \bibinfo{booktitle}{\emph{Proceedings of the 53rd Annual
  Meeting of the Association for Computational Linguistics and the 7th
  International Joint Conference on Natural Language Processing (Volume 2:
  Short Papers)}}. \bibinfo{pages}{438--444}.
\newblock


\bibitem[Chang et~al\mbox{.}(2009)]%
        {chang2009reading}
\bibfield{author}{\bibinfo{person}{Jonathan Chang}, \bibinfo{person}{Sean
  Gerrish}, \bibinfo{person}{Chong Wang}, \bibinfo{person}{Jordan Boyd-Graber},
  {and} \bibinfo{person}{David Blei}.} \bibinfo{year}{2009}\natexlab{}.
\newblock \showarticletitle{Reading tea leaves: How humans interpret topic
  models}.
\newblock \bibinfo{journal}{\emph{Advances in neural information processing
  systems}}  \bibinfo{volume}{22} (\bibinfo{year}{2009}).
\newblock


\bibitem[Chen et~al\mbox{.}(2022a)]%
        {chen2022semeval}
\bibfield{author}{\bibinfo{person}{Xi Chen}, \bibinfo{person}{Ali Zeynali},
  \bibinfo{person}{Chico Camargo}, \bibinfo{person}{Fabian Fl{\"o}ck},
  \bibinfo{person}{Devin Gaffney}, \bibinfo{person}{Przemyslaw Grabowicz},
  \bibinfo{person}{Scott Hale}, \bibinfo{person}{David Jurgens}, {and}
  \bibinfo{person}{Mattia Samory}.} \bibinfo{year}{2022}\natexlab{a}.
\newblock \showarticletitle{SemEval-2022 Task 8: Multilingual news article
  similarity}. In \bibinfo{booktitle}{\emph{SemEval-2022}}.
  \bibinfo{pages}{1094--1106}.
\newblock


\bibitem[Chen et~al\mbox{.}(2022b)]%
        {chen-etal-2022-semeval}
\bibfield{author}{\bibinfo{person}{Xi Chen}, \bibinfo{person}{Ali Zeynali},
  \bibinfo{person}{Chico Camargo}, \bibinfo{person}{Fabian Fl{\"o}ck},
  \bibinfo{person}{Devin Gaffney}, \bibinfo{person}{Przemyslaw Grabowicz},
  \bibinfo{person}{Scott Hale}, \bibinfo{person}{David Jurgens}, {and}
  \bibinfo{person}{Mattia Samory}.} \bibinfo{year}{2022}\natexlab{b}.
\newblock \showarticletitle{{S}em{E}val-2022 Task 8: Multilingual news article
  similarity}. In \bibinfo{booktitle}{\emph{SemEval-2022}}.
  \bibinfo{publisher}{Association for Computational Linguistics},
  \bibinfo{address}{Seattle, United States}, \bibinfo{pages}{1094--1106}.
\newblock
\urldef\tempurl%
\url{https://doi.org/10.18653/v1/2022.semeval-1.155}
\showDOI{\tempurl}


\bibitem[Choi(2009)]%
        {Choi2009-pc}
\bibfield{author}{\bibinfo{person}{Jihyang Choi}.}
  \bibinfo{year}{2009}\natexlab{}.
\newblock \showarticletitle{Diversity in Foreign News in {US} Newspapers Before
  and After the Invasion of Iraq}.
\newblock \bibinfo{journal}{\emph{International Communication Gazette}}
  \bibinfo{volume}{71}, \bibinfo{number}{6} (\bibinfo{date}{Oct.}
  \bibinfo{year}{2009}), \bibinfo{pages}{525--542}.
\newblock


\bibitem[Cohen(1963)]%
        {cohen1963press}
\bibfield{author}{\bibinfo{person}{Bernard~C. Cohen}.}
  \bibinfo{year}{1963}\natexlab{}.
\newblock \bibinfo{booktitle}{\emph{Press and Foreign Policy}}.
\newblock \bibinfo{publisher}{Princeton University Press}.
\newblock
\showISBNx{9780691075198}
\urldef\tempurl%
\url{http://www.jstor.org/stable/j.ctt183q0fp}
\showURL{%
\tempurl}


\bibitem[Conneau et~al\mbox{.}(2017)]%
        {conneau2017word}
\bibfield{author}{\bibinfo{person}{Alexis Conneau}, \bibinfo{person}{Guillaume
  Lample}, \bibinfo{person}{Marc'Aurelio Ranzato}, \bibinfo{person}{Ludovic
  Denoyer}, {and} \bibinfo{person}{Herv{\'e} J{\'e}gou}.}
  \bibinfo{year}{2017}\natexlab{}.
\newblock \showarticletitle{Word Translation Without Parallel Data}.
\newblock \bibinfo{journal}{\emph{arXiv preprint arXiv:1710.04087}}
  (\bibinfo{year}{2017}).
\newblock


\bibitem[Dupree(1971)]%
        {dupree1971international}
\bibfield{author}{\bibinfo{person}{John~David Dupree}.}
  \bibinfo{year}{1971}\natexlab{}.
\newblock \showarticletitle{International communication: View from'a window on
  the world'}.
\newblock \bibinfo{journal}{\emph{Gazette (Leiden, Netherlands)}}
  \bibinfo{volume}{17}, \bibinfo{number}{4} (\bibinfo{year}{1971}),
  \bibinfo{pages}{224--235}.
\newblock


\bibitem[Eisensee and Str{\"o}mberg(2007)]%
        {eisensee2007news}
\bibfield{author}{\bibinfo{person}{Thomas Eisensee} {and}
  \bibinfo{person}{David Str{\"o}mberg}.} \bibinfo{year}{2007}\natexlab{}.
\newblock \showarticletitle{News droughts, news floods, and US disaster
  relief}.
\newblock \bibinfo{journal}{\emph{The Quarterly Journal of Economics}}
  \bibinfo{volume}{122}, \bibinfo{number}{2} (\bibinfo{year}{2007}),
  \bibinfo{pages}{693--728}.
\newblock


\bibitem[Feng et~al\mbox{.}(2020a)]%
        {Feng2020-cw}
\bibfield{author}{\bibinfo{person}{Chong Feng}, \bibinfo{person}{Muzammil
  Khan}, \bibinfo{person}{Arif~Ur Rahman}, {and} \bibinfo{person}{Arshad
  Ahmad}.} \bibinfo{year}{2020}\natexlab{a}.
\newblock \showarticletitle{News Recommendation Systems - Accomplishments,
  Challenges \& Future Directions}.
\newblock \bibinfo{journal}{\emph{IEEE Access}}  \bibinfo{volume}{8}
  (\bibinfo{year}{2020}), \bibinfo{pages}{16702--16725}.
\newblock


\bibitem[Feng et~al\mbox{.}(2020b)]%
        {feng2020language}
\bibfield{author}{\bibinfo{person}{Fangxiaoyu Feng}, \bibinfo{person}{Yinfei
  Yang}, \bibinfo{person}{Daniel Cer}, \bibinfo{person}{Naveen Arivazhagan},
  {and} \bibinfo{person}{Wei Wang}.} \bibinfo{year}{2020}\natexlab{b}.
\newblock \showarticletitle{Language-agnostic BERT sentence embedding}.
\newblock \bibinfo{journal}{\emph{arXiv preprint arXiv:2007.01852}}
  (\bibinfo{year}{2020}).
\newblock


\bibitem[Feng et~al\mbox{.}(2022)]%
        {feng-etal-2022-language}
\bibfield{author}{\bibinfo{person}{Fangxiaoyu Feng}, \bibinfo{person}{Yinfei
  Yang}, \bibinfo{person}{Daniel Cer}, \bibinfo{person}{Naveen Arivazhagan},
  {and} \bibinfo{person}{Wei Wang}.} \bibinfo{year}{2022}\natexlab{}.
\newblock \showarticletitle{Language-agnostic {BERT} Sentence Embedding}. In
  \bibinfo{booktitle}{\emph{Proceedings of the 60th Annual Meeting of the
  Association for Computational Linguistics (Volume 1: Long Papers)}}.
  \bibinfo{publisher}{Association for Computational Linguistics},
  \bibinfo{address}{Dublin, Ireland}, \bibinfo{pages}{878--891}.
\newblock
\urldef\tempurl%
\url{https://doi.org/10.18653/v1/2022.acl-long.62}
\showDOI{\tempurl}


\bibitem[Fuglede and Topsoe(2004)]%
        {fuglede2004jensen}
\bibfield{author}{\bibinfo{person}{Bent Fuglede} {and}
  \bibinfo{person}{Flemming Topsoe}.} \bibinfo{year}{2004}\natexlab{}.
\newblock \showarticletitle{Jensen-Shannon divergence and Hilbert space
  embedding}. In \bibinfo{booktitle}{\emph{International Symposium on
  Information Theory, 2004. ISIT 2004. Proceedings.}} IEEE,
  \bibinfo{pages}{31}.
\newblock


\bibitem[Galtung and Ruge(1965)]%
        {galtung1965structure}
\bibfield{author}{\bibinfo{person}{Johan Galtung} {and}
  \bibinfo{person}{Mari~Holmboe Ruge}.} \bibinfo{year}{1965}\natexlab{}.
\newblock \showarticletitle{The structure of foreign news: The presentation of
  the Congo, Cuba and Cyprus crises in four Norwegian newspapers}.
\newblock \bibinfo{journal}{\emph{Journal of peace research}}
  \bibinfo{volume}{2}, \bibinfo{number}{1} (\bibinfo{year}{1965}),
  \bibinfo{pages}{64--90}.
\newblock


\bibitem[Gerbner and Marvanyi(1977)]%
        {gerbner1977many}
\bibfield{author}{\bibinfo{person}{George Gerbner} {and}
  \bibinfo{person}{George Marvanyi}.} \bibinfo{year}{1977}\natexlab{}.
\newblock \showarticletitle{The many worlds of the world's press}.
\newblock \bibinfo{journal}{\emph{Journal of communication}}
  \bibinfo{volume}{27}, \bibinfo{number}{1} (\bibinfo{year}{1977}),
  \bibinfo{pages}{52--66}.
\newblock


\bibitem[Golan and Himelboim(2016)]%
        {golan2016can}
\bibfield{author}{\bibinfo{person}{Guy~J Golan} {and} \bibinfo{person}{Itai
  Himelboim}.} \bibinfo{year}{2016}\natexlab{}.
\newblock \showarticletitle{Can World System Theory predict news flow on
  twitter? The case of government-sponsored broadcasting}.
\newblock \bibinfo{journal}{\emph{Information, Communication \& Society}}
  \bibinfo{volume}{19}, \bibinfo{number}{8} (\bibinfo{year}{2016}),
  \bibinfo{pages}{1150--1170}.
\newblock


\bibitem[Grabowicz et~al\mbox{.}(2012)]%
        {grabowicz2012social}
\bibfield{author}{\bibinfo{person}{Przemyslaw~A. Grabowicz},
  \bibinfo{person}{Jos{\'e}~J. Ramasco}, \bibinfo{person}{Esteban Moro},
  \bibinfo{person}{Josep~M. Pujol}, {and} \bibinfo{person}{Victor~M. Eguiluz}.}
  \bibinfo{year}{2012}\natexlab{}.
\newblock \showarticletitle{Social {{Features}} of {{Online Networks}}: {{The
  Strength}} of {{Intermediary Ties}} in {{Online Social Media}}}.
\newblock \bibinfo{journal}{\emph{PLoS ONE}} \bibinfo{volume}{7},
  \bibinfo{number}{1} (\bibinfo{date}{Jan.} \bibinfo{year}{2012}),
  \bibinfo{pages}{e29358}.
\newblock
\showISSN{1932-6203}
\urldef\tempurl%
\url{https://doi.org/10.1371/journal.pone.0029358}
\showDOI{\tempurl}


\bibitem[Grasland(2020)]%
        {grasland2020international}
\bibfield{author}{\bibinfo{person}{Claude Grasland}.}
  \bibinfo{year}{2020}\natexlab{}.
\newblock \showarticletitle{International news flow theory revisited through a
  space--time interaction model: Application to a sample of 320,000
  international news stories published through RSS flows by 31 daily newspapers
  in 2015}.
\newblock \bibinfo{journal}{\emph{International Communication Gazette}}
  \bibinfo{volume}{82}, \bibinfo{number}{3} (\bibinfo{year}{2020}),
  \bibinfo{pages}{231--259}.
\newblock


\bibitem[Greer(2003)]%
        {greer2003sex}
\bibfield{author}{\bibinfo{person}{Chris Greer}.}
  \bibinfo{year}{2003}\natexlab{}.
\newblock \bibinfo{booktitle}{\emph{Sex Crime and the Media: Sex Offending and
  the Press in a Divided Society}}.
\newblock \bibinfo{publisher}{Willan}.
\newblock
\showISBNx{9781843924869}
\urldef\tempurl%
\url{https://doi.org/10.4324/9781843924869}
\showURL{%
\tempurl}


\bibitem[Guo and Vargo(2017)]%
        {guo2017global}
\bibfield{author}{\bibinfo{person}{Lei Guo} {and} \bibinfo{person}{Chris~J
  Vargo}.} \bibinfo{year}{2017}\natexlab{}.
\newblock \showarticletitle{Global intermedia agenda setting: A big data
  analysis of international news flow}.
\newblock \bibinfo{journal}{\emph{Journal of Communication}}
  \bibinfo{volume}{67}, \bibinfo{number}{4} (\bibinfo{year}{2017}),
  \bibinfo{pages}{499--520}.
\newblock


\bibitem[Guo and Vargo(2020)]%
        {guo2020predictors}
\bibfield{author}{\bibinfo{person}{Lei Guo} {and} \bibinfo{person}{Chris~J
  Vargo}.} \bibinfo{year}{2020}\natexlab{}.
\newblock \showarticletitle{Predictors of international news flow: Exploring a
  networked global media system}.
\newblock \bibinfo{journal}{\emph{Journal of Broadcasting \& Electronic Media}}
  \bibinfo{volume}{64}, \bibinfo{number}{3} (\bibinfo{year}{2020}),
  \bibinfo{pages}{418--437}.
\newblock


\bibitem[Gusev et~al\mbox{.}(2021)]%
        {Gusev2021-tc}
\bibfield{author}{\bibinfo{person}{Ilya Gusev}, \bibinfo{person}{{Moscow
  Institute of Physics and Technology}}, \bibinfo{person}{Ivan Smurov}, {and}
  \bibinfo{person}{{ABBYY}}.} \bibinfo{year}{2021}\natexlab{}.
\newblock \bibinfo{title}{Russian News Clustering and Headline Selection Shared
  Task}.
\newblock
\newblock


\bibitem[Hester(1973)]%
        {hester1973theoretical}
\bibfield{author}{\bibinfo{person}{Albert~L Hester}.}
  \bibinfo{year}{1973}\natexlab{}.
\newblock \showarticletitle{Theoretical considerations in predicting volume and
  direction of international information flow}.
\newblock \bibinfo{journal}{\emph{Gazette (Leiden, Netherlands)}}
  \bibinfo{volume}{19}, \bibinfo{number}{4} (\bibinfo{year}{1973}),
  \bibinfo{pages}{239--247}.
\newblock


\bibitem[Huang et~al\mbox{.}(2016)]%
        {huang2016liberal}
\bibfield{author}{\bibinfo{person}{Lifu Huang}, \bibinfo{person}{Taylor
  Cassidy}, \bibinfo{person}{Xiaocheng Feng}, \bibinfo{person}{Heng Ji},
  \bibinfo{person}{Clare Voss}, \bibinfo{person}{Jiawei Han}, {and}
  \bibinfo{person}{Avirup Sil}.} \bibinfo{year}{2016}\natexlab{}.
\newblock \showarticletitle{Liberal event extraction and event schema
  induction}. In \bibinfo{booktitle}{\emph{Proceedings of the 54th Annual
  Meeting of the Association for Computational Linguistics (Volume 1: Long
  Papers)}}. \bibinfo{pages}{258--268}.
\newblock


\bibitem[Hughes and Prado(2011)]%
        {hughes2011media}
\bibfield{author}{\bibinfo{person}{Sallie Hughes} {and} \bibinfo{person}{Paola
  Prado}.} \bibinfo{year}{2011}\natexlab{}.
\newblock \showarticletitle{Media diversity and social inequality in Latin
  America}.
\newblock \bibinfo{journal}{\emph{The great gap: Inequality and the politics of
  redistribution in Latin America}} (\bibinfo{year}{2011}),
  \bibinfo{pages}{109--146}.
\newblock


\bibitem[Jansen et~al\mbox{.}(2019)]%
        {jansen2019drives}
\bibfield{author}{\bibinfo{person}{A~Severin Jansen}, \bibinfo{person}{Beatrice
  Eugster}, \bibinfo{person}{Michaela Maier}, {and} \bibinfo{person}{Silke
  Adam}.} \bibinfo{year}{2019}\natexlab{}.
\newblock \showarticletitle{Who drives the agenda: Media or parties? A
  seven-country comparison in the run-up to the 2014 European Parliament
  elections}.
\newblock \bibinfo{journal}{\emph{The International Journal of Press/Politics}}
  \bibinfo{volume}{24}, \bibinfo{number}{1} (\bibinfo{year}{2019}),
  \bibinfo{pages}{7--26}.
\newblock


\bibitem[Journalism.org(2016)]%
        {noauthor2016-gb}
\bibfield{author}{\bibinfo{person}{Journalism.org}.}
  \bibinfo{year}{2016}\natexlab{}.
\newblock \bibinfo{title}{The state of the news media}.
\newblock
  \bibinfo{howpublished}{\url{https://assets.pewresearch.org/wp-content/uploads/sites/13/2016/06/30143308/state-of-the-news-media-report-2016-final.pdf}}.
\newblock
\newblock
\shownote{Accessed: 2023-12-1}.


\bibitem[Kariel and Rosenvall(1984)]%
        {kariel1984factors}
\bibfield{author}{\bibinfo{person}{Herbert~G Kariel} {and}
  \bibinfo{person}{Lynn~A Rosenvall}.} \bibinfo{year}{1984}\natexlab{}.
\newblock \showarticletitle{Factors influencing international news flow}.
\newblock \bibinfo{journal}{\emph{Journalism Quarterly}} \bibinfo{volume}{61},
  \bibinfo{number}{3} (\bibinfo{year}{1984}), \bibinfo{pages}{509--666}.
\newblock


\bibitem[Kim and Barnett(1996)]%
        {kim1996determinants}
\bibfield{author}{\bibinfo{person}{Kyungmo Kim} {and} \bibinfo{person}{George~A
  Barnett}.} \bibinfo{year}{1996}\natexlab{}.
\newblock \showarticletitle{The determinants of international news flow: A
  network analysis}.
\newblock \bibinfo{journal}{\emph{Communication Research}}
  \bibinfo{volume}{23}, \bibinfo{number}{3} (\bibinfo{year}{1996}),
  \bibinfo{pages}{323--352}.
\newblock


\bibitem[Klinenberg(2005)]%
        {Klinenberg2005-ol}
\bibfield{author}{\bibinfo{person}{Eric Klinenberg}.}
  \bibinfo{year}{2005}\natexlab{}.
\newblock \showarticletitle{Convergence: News Production in a Digital Age}.
\newblock \bibinfo{journal}{\emph{Ann. Am. Acad. Pol. Soc. Sci.}}
  \bibinfo{volume}{597}, \bibinfo{number}{1} (\bibinfo{date}{Jan.}
  \bibinfo{year}{2005}), \bibinfo{pages}{48--64}.
\newblock


\bibitem[Kumaran and Allan(2004)]%
        {kumaran2004text}
\bibfield{author}{\bibinfo{person}{Giridhar Kumaran} {and}
  \bibinfo{person}{James Allan}.} \bibinfo{year}{2004}\natexlab{}.
\newblock \showarticletitle{Text classification and named entities for new
  event detection}. In \bibinfo{booktitle}{\emph{Proceedings of the 27th annual
  international ACM SIGIR conference on Research and development in information
  retrieval}}. \bibinfo{pages}{297--304}.
\newblock


\bibitem[Lam et~al\mbox{.}(2001)]%
        {lam2001using}
\bibfield{author}{\bibinfo{person}{Wai Lam}, \bibinfo{person}{HML Meng},
  \bibinfo{person}{KL Wong}, {and} \bibinfo{person}{JCH Yen}.}
  \bibinfo{year}{2001}\natexlab{}.
\newblock \showarticletitle{Using contextual analysis for news event
  detection}.
\newblock \bibinfo{journal}{\emph{International Journal of Intelligent
  Systems}} \bibinfo{volume}{16}, \bibinfo{number}{4} (\bibinfo{year}{2001}),
  \bibinfo{pages}{525--546}.
\newblock


\bibitem[Lancichinetti et~al\mbox{.}(2011)]%
        {lancichinetti2011finding}
\bibfield{author}{\bibinfo{person}{Andrea Lancichinetti},
  \bibinfo{person}{Filippo Radicchi}, \bibinfo{person}{Jos{\'e}~J Ramasco},
  {and} \bibinfo{person}{Santo Fortunato}.} \bibinfo{year}{2011}\natexlab{}.
\newblock \showarticletitle{Finding statistically significant communities in
  networks}.
\newblock \bibinfo{journal}{\emph{PloS one}} \bibinfo{volume}{6},
  \bibinfo{number}{4} (\bibinfo{year}{2011}), \bibinfo{pages}{e18961}.
\newblock


\bibitem[Lee(2013)]%
        {lee2013news}
\bibfield{author}{\bibinfo{person}{Angela~M Lee}.}
  \bibinfo{year}{2013}\natexlab{}.
\newblock \showarticletitle{News audiences revisited: Theorizing the link
  between audience motivations and news consumption}.
\newblock \bibinfo{journal}{\emph{Journal of Broadcasting \& Electronic Media}}
  \bibinfo{volume}{57}, \bibinfo{number}{3} (\bibinfo{year}{2013}),
  \bibinfo{pages}{300--317}.
\newblock


\bibitem[Lee(2007)]%
        {lee2007international}
\bibfield{author}{\bibinfo{person}{Suman Lee}.}
  \bibinfo{year}{2007}\natexlab{}.
\newblock \showarticletitle{International public relations as a predictor of
  prominence of US news coverage}.
\newblock \bibinfo{journal}{\emph{Public Relations Review}}
  \bibinfo{volume}{33}, \bibinfo{number}{2} (\bibinfo{year}{2007}),
  \bibinfo{pages}{158--165}.
\newblock


\bibitem[Leetaru and Schrodt(2013)]%
        {leetaru2013gdelt}
\bibfield{author}{\bibinfo{person}{Kalev Leetaru} {and}
  \bibinfo{person}{Philip~A Schrodt}.} \bibinfo{year}{2013}\natexlab{}.
\newblock \showarticletitle{Gdelt: Global data on events, location, and tone,
  1979--2012}. In \bibinfo{booktitle}{\emph{ISA annual convention}},
  Vol.~\bibinfo{volume}{2}. Citeseer, \bibinfo{pages}{1--49}.
\newblock


\bibitem[Leskovec et~al\mbox{.}(2009)]%
        {Leskovec2009-ob}
\bibfield{author}{\bibinfo{person}{Jure Leskovec}, \bibinfo{person}{Lars
  Backstrom}, {and} \bibinfo{person}{Jon Kleinberg}.}
  \bibinfo{year}{2009}\natexlab{}.
\newblock \showarticletitle{Meme-tracking and the dynamics of the news cycle}.
  In \bibinfo{booktitle}{\emph{{KDD}}} (Paris, France)
  \emph{(\bibinfo{series}{KDD '09})}. \bibinfo{pages}{497--506}.
\newblock


\bibitem[Li et~al\mbox{.}(2005)]%
        {li2005probabilistic}
\bibfield{author}{\bibinfo{person}{Zhiwei Li}, \bibinfo{person}{Bin Wang},
  \bibinfo{person}{Mingjing Li}, {and} \bibinfo{person}{Wei-Ying Ma}.}
  \bibinfo{year}{2005}\natexlab{}.
\newblock \showarticletitle{A probabilistic model for retrospective news event
  detection}. In \bibinfo{booktitle}{\emph{Proceedings of the 28th annual
  international ACM SIGIR conference on Research and development in information
  retrieval}}. \bibinfo{pages}{106--113}.
\newblock


\bibitem[Litterer et~al\mbox{.}(2023)]%
        {litterer2023rains}
\bibfield{author}{\bibinfo{person}{Benjamin Litterer}, \bibinfo{person}{David
  Jurgens}, {and} \bibinfo{person}{Dallas Card}.}
  \bibinfo{year}{2023}\natexlab{}.
\newblock \showarticletitle{When it Rains, it Pours: Modeling Media Storms and
  the News Ecosystem}. In \bibinfo{booktitle}{\emph{Findings of the Association
  for Computational Linguistics: EMNLP 2023}}. \bibinfo{pages}{6346--6361}.
\newblock


\bibitem[McCombs and Reynolds(2002)]%
        {mccombs2002news}
\bibfield{author}{\bibinfo{person}{Maxwell McCombs} {and} \bibinfo{person}{Amy
  Reynolds}.} \bibinfo{year}{2002}\natexlab{}.
\newblock \showarticletitle{News influence on our pictures of the world}.
\newblock In \bibinfo{booktitle}{\emph{Media effects}}.
  \bibinfo{publisher}{Routledge}, \bibinfo{pages}{11--28}.
\newblock


\bibitem[McCombs and Valenzuela(2021)]%
        {mccombs2021setting}
\bibfield{author}{\bibinfo{person}{Maxwell McCombs} {and}
  \bibinfo{person}{Sebastian Valenzuela}.} \bibinfo{year}{2021}\natexlab{}.
\newblock \bibinfo{booktitle}{\emph{Setting the Agenda: Mass Media and Public
  Opinion}}.
\newblock \bibinfo{publisher}{Wiley}.
\newblock


\bibitem[McCombs and Shaw(1972)]%
        {mccombs1972agenda}
\bibfield{author}{\bibinfo{person}{Maxwell~E McCombs} {and}
  \bibinfo{person}{Donald~L Shaw}.} \bibinfo{year}{1972}\natexlab{}.
\newblock \showarticletitle{The agenda-setting function of mass media}.
\newblock \bibinfo{journal}{\emph{Public opinion quarterly}}
  \bibinfo{volume}{36}, \bibinfo{number}{2} (\bibinfo{year}{1972}),
  \bibinfo{pages}{176--187}.
\newblock


\bibitem[McCombs and Shaw(1993)]%
        {mccombs1993evolution}
\bibfield{author}{\bibinfo{person}{Maxwell~E McCombs} {and}
  \bibinfo{person}{Donald~L Shaw}.} \bibinfo{year}{1993}\natexlab{}.
\newblock \showarticletitle{The evolution of agenda-setting research:
  Twenty-five years in the marketplace of ideas}.
\newblock \bibinfo{journal}{\emph{Journal of communication}}
  \bibinfo{volume}{43}, \bibinfo{number}{2} (\bibinfo{year}{1993}),
  \bibinfo{pages}{58--67}.
\newblock


\bibitem[McGregor(2019)]%
        {mcgregor2019social}
\bibfield{author}{\bibinfo{person}{Shannon~C McGregor}.}
  \bibinfo{year}{2019}\natexlab{}.
\newblock \showarticletitle{Social media as public opinion: How journalists use
  social media to represent public opinion}.
\newblock \bibinfo{journal}{\emph{Journalism}} \bibinfo{volume}{20},
  \bibinfo{number}{8} (\bibinfo{year}{2019}), \bibinfo{pages}{1070--1086}.
\newblock


\bibitem[Nicholls(2019)]%
        {Nicholls2019-gu}
\bibfield{author}{\bibinfo{person}{T Nicholls}.}
  \bibinfo{year}{2019}\natexlab{}.
\newblock \showarticletitle{Detecting textual reuse in news stories, at scale}.
\newblock \bibinfo{journal}{\emph{Int. J. Commun. Syst.}} \bibinfo{volume}{13},
  \bibinfo{number}{2019} (\bibinfo{year}{2019}).
\newblock


\bibitem[Nnaemeka and Richstad(1981)]%
        {nnaemeka1981internal}
\bibfield{author}{\bibinfo{person}{Tony Nnaemeka} {and} \bibinfo{person}{Jim
  Richstad}.} \bibinfo{year}{1981}\natexlab{}.
\newblock \showarticletitle{Internal Controls and Foreign News Coverage:
  Pacific Press Systems}.
\newblock \bibinfo{journal}{\emph{Communication Research}} \bibinfo{volume}{8},
  \bibinfo{number}{1} (\bibinfo{year}{1981}), \bibinfo{pages}{97--135}.
\newblock


\bibitem[Page et~al\mbox{.}(1999)]%
        {page1999pagerank}
\bibfield{author}{\bibinfo{person}{Lawrence Page}, \bibinfo{person}{Sergey
  Brin}, \bibinfo{person}{Rajeev Motwani}, {and} \bibinfo{person}{Terry
  Winograd}.} \bibinfo{year}{1999}\natexlab{}.
\newblock \bibinfo{booktitle}{\emph{The PageRank citation ranking: Bringing
  order to the web.}}
\newblock \bibinfo{type}{{T}echnical {R}eport}. \bibinfo{institution}{Stanford
  InfoLab}.
\newblock


\bibitem[Paulussen and Van~Aelst(2021)]%
        {paulussen2021news}
\bibfield{author}{\bibinfo{person}{Steve Paulussen} {and}
  \bibinfo{person}{Peter Van~Aelst}.} \bibinfo{year}{2021}\natexlab{}.
\newblock \showarticletitle{News values in audience-oriented journalism:
  Criteria, angles, and cues of Newsworthiness in the (Digital) media context}.
\newblock \bibinfo{journal}{\emph{News values from an audience perspective}}
  (\bibinfo{year}{2021}), \bibinfo{pages}{37--55}.
\newblock


\bibitem[Po{\"{}}~ttker(2003)]%
        {po2003news}
\bibfield{author}{\bibinfo{person}{Horst Po{\"{}}~ttker}.}
  \bibinfo{year}{2003}\natexlab{}.
\newblock \showarticletitle{News and its communicative quality: the inverted
  pyramid—when and why did it appear?}
\newblock \bibinfo{journal}{\emph{Journalism Studies}} \bibinfo{volume}{4},
  \bibinfo{number}{4} (\bibinfo{year}{2003}), \bibinfo{pages}{501--511}.
\newblock


\bibitem[Pranjic et~al\mbox{.}(2020)]%
        {Pranjic2020-st}
\bibfield{author}{\bibinfo{person}{Marko Pranjic}, \bibinfo{person}{Vid
  Podpecan}, \bibinfo{person}{Marko Robnik-{\v S}ikonja}, {and}
  \bibinfo{person}{Senja Pollak}.} \bibinfo{year}{2020}\natexlab{}.
\newblock \showarticletitle{Evaluation of related news recommendations using
  document similarity methods}. In \bibinfo{booktitle}{\emph{Proceedings of the
  Conference on Language Technologies and Digital Humanities, Ljubljana}}.
  \bibinfo{publisher}{nl.ijs.si}, \bibinfo{pages}{81--86}.
\newblock


\bibitem[Reimers and Gurevych(2019)]%
        {reimers-2019-sentence-bert}
\bibfield{author}{\bibinfo{person}{Nils Reimers} {and} \bibinfo{person}{Iryna
  Gurevych}.} \bibinfo{year}{2019}\natexlab{}.
\newblock \showarticletitle{Sentence-BERT: Sentence Embeddings using Siamese
  BERT-Networks}. In \bibinfo{booktitle}{\emph{EMNLP}}.
  \bibinfo{publisher}{Association for Computational Linguistics}.
\newblock
\urldef\tempurl%
\url{https://arxiv.org/abs/1908.10084}
\showURL{%
\tempurl}


\bibitem[Roberts et~al\mbox{.}(2021)]%
        {roberts2021media}
\bibfield{author}{\bibinfo{person}{Hal Roberts}, \bibinfo{person}{Rahul
  Bhargava}, \bibinfo{person}{Linas Valiukas}, \bibinfo{person}{Dennis Jen},
  \bibinfo{person}{Momin~M Malik}, \bibinfo{person}{Cindy~Sherman Bishop},
  \bibinfo{person}{Emily~B Ndulue}, \bibinfo{person}{Aashka Dave},
  \bibinfo{person}{Justin Clark}, \bibinfo{person}{Bruce Etling},
  {et~al\mbox{.}}} \bibinfo{year}{2021}\natexlab{}.
\newblock \showarticletitle{Media cloud: Massive open source collection of
  global news on the open web}.
\newblock \bibinfo{journal}{\emph{ICWSM}} (\bibinfo{year}{2021}).
\newblock


\bibitem[Robinson and Sparkes(1976)]%
        {robinson1976international}
\bibfield{author}{\bibinfo{person}{Gertrude~Joch Robinson} {and}
  \bibinfo{person}{Vernone~M Sparkes}.} \bibinfo{year}{1976}\natexlab{}.
\newblock \showarticletitle{International news in the Canadian and American
  press: A comparative news flow study}.
\newblock \bibinfo{journal}{\emph{Gazette (Leiden, Netherlands)}}
  \bibinfo{volume}{22}, \bibinfo{number}{4} (\bibinfo{year}{1976}),
  \bibinfo{pages}{203--218}.
\newblock


\bibitem[Scollon(2000)]%
        {Scollon2000-bi}
\bibfield{author}{\bibinfo{person}{Ron Scollon}.}
  \bibinfo{year}{2000}\natexlab{}.
\newblock \showarticletitle{Generic variability in news stories in Chinese and
  English: A contrastive discourse study of five days' newspapers}.
\newblock \bibinfo{journal}{\emph{J. Pragmat.}} \bibinfo{volume}{32},
  \bibinfo{number}{6} (\bibinfo{date}{May} \bibinfo{year}{2000}),
  \bibinfo{pages}{761--791}.
\newblock


\bibitem[Segev(2015)]%
        {segev2015visible}
\bibfield{author}{\bibinfo{person}{Elad Segev}.}
  \bibinfo{year}{2015}\natexlab{}.
\newblock \showarticletitle{Visible and invisible countries: News flow theory
  revised}.
\newblock \bibinfo{journal}{\emph{Journalism}} \bibinfo{volume}{16},
  \bibinfo{number}{3} (\bibinfo{year}{2015}), \bibinfo{pages}{412--428}.
\newblock


\bibitem[Segev(2016)]%
        {segev2016group}
\bibfield{author}{\bibinfo{person}{Elad Segev}.}
  \bibinfo{year}{2016}\natexlab{}.
\newblock \showarticletitle{The group-sphere model of international news flow:
  A cross-national comparison of news sites}.
\newblock \bibinfo{journal}{\emph{International Communication Gazette}}
  \bibinfo{volume}{78}, \bibinfo{number}{3} (\bibinfo{year}{2016}),
  \bibinfo{pages}{200--222}.
\newblock


\bibitem[Segev and Blondheim(2013)]%
        {segev2013america}
\bibfield{author}{\bibinfo{person}{Elad Segev} {and} \bibinfo{person}{Menahem
  Blondheim}.} \bibinfo{year}{2013}\natexlab{}.
\newblock \showarticletitle{America's global standing according to popular news
  sites from around the world}.
\newblock \bibinfo{journal}{\emph{Political Communication}}
  \bibinfo{volume}{30}, \bibinfo{number}{1} (\bibinfo{year}{2013}),
  \bibinfo{pages}{139--161}.
\newblock


\bibitem[Semmel(1977)]%
        {semmel1977elite}
\bibfield{author}{\bibinfo{person}{Andrew~K. Semmel}.}
  \bibinfo{year}{1977}\natexlab{}.
\newblock \showarticletitle{The elite press, the global system, and foreign
  news attention}.
\newblock \bibinfo{journal}{\emph{International Interactions}}
  \bibinfo{volume}{3}, \bibinfo{number}{4} (\bibinfo{year}{1977}),
  \bibinfo{pages}{317--328}.
\newblock
\urldef\tempurl%
\url{https://doi.org/10.1080/03050627708434471}
\showDOI{\tempurl}


\bibitem[Serrano et~al\mbox{.}(2009)]%
        {serrano2009extracting}
\bibfield{author}{\bibinfo{person}{M~{\'A}ngeles Serrano},
  \bibinfo{person}{Mari{\'a}n Bogun{\'a}}, {and} \bibinfo{person}{Alessandro
  Vespignani}.} \bibinfo{year}{2009}\natexlab{}.
\newblock \showarticletitle{Extracting the multiscale backbone of complex
  weighted networks}.
\newblock \bibinfo{journal}{\emph{PNAS}} \bibinfo{volume}{106},
  \bibinfo{number}{16} (\bibinfo{year}{2009}), \bibinfo{pages}{6483--6488}.
\newblock


\bibitem[Shannon(1948)]%
        {shannon1948mathematical}
\bibfield{author}{\bibinfo{person}{Claude~Elwood Shannon}.}
  \bibinfo{year}{1948}\natexlab{}.
\newblock \showarticletitle{A mathematical theory of communication}.
\newblock \bibinfo{journal}{\emph{The Bell system technical journal}}
  \bibinfo{volume}{27}, \bibinfo{number}{3} (\bibinfo{year}{1948}),
  \bibinfo{pages}{379--423}.
\newblock


\bibitem[Shoemaker and Reese(1996)]%
        {Shoemaker1996-gp}
\bibfield{author}{\bibinfo{person}{Pamela~J Shoemaker} {and}
  \bibinfo{person}{Stephen~D Reese}.} \bibinfo{year}{1996}\natexlab{}.
\newblock \bibinfo{booktitle}{\emph{Mediating the Message: Theories of
  Influences on Mass Media Content}}.
\newblock \bibinfo{publisher}{Longman}.
\newblock


\bibitem[Shoemaker and Vos(2009)]%
        {shoemaker2009gatekeeping}
\bibfield{author}{\bibinfo{person}{Pamela~J Shoemaker} {and}
  \bibinfo{person}{Timothy Vos}.} \bibinfo{year}{2009}\natexlab{}.
\newblock \bibinfo{booktitle}{\emph{Gatekeeping theory}}.
\newblock \bibinfo{publisher}{Routledge}.
\newblock


\bibitem[Singh et~al\mbox{.}(2022)]%
        {singh-etal-2022-gatenlp}
\bibfield{author}{\bibinfo{person}{Iknoor Singh}, \bibinfo{person}{Yue Li},
  \bibinfo{person}{Melissa Thong}, {and} \bibinfo{person}{Carolina Scarton}.}
  \bibinfo{year}{2022}\natexlab{}.
\newblock \showarticletitle{{G}ate{NLP}-{US}hef at {S}em{E}val-2022 Task 8:
  Entity-Enriched {S}iamese Transformer for Multilingual News Article
  Similarity}. In \bibinfo{booktitle}{\emph{SemEval-2022}}.
  \bibinfo{publisher}{Association for Computational Linguistics},
  \bibinfo{address}{Seattle, United States}, \bibinfo{pages}{1121--1128}.
\newblock
\urldef\tempurl%
\url{https://doi.org/10.18653/v1/2022.semeval-1.158}
\showDOI{\tempurl}


\bibitem[Song et~al\mbox{.}(2020)]%
        {Song2020MPNetMA}
\bibfield{author}{\bibinfo{person}{Kaitao Song}, \bibinfo{person}{Xu Tan},
  \bibinfo{person}{Tao Qin}, \bibinfo{person}{Jianfeng Lu}, {and}
  \bibinfo{person}{Tie-Yan Liu}.} \bibinfo{year}{2020}\natexlab{}.
\newblock \showarticletitle{MPNet: Masked and Permuted Pre-training for
  Language Understanding}.
\newblock \bibinfo{journal}{\emph{ArXiv}}  \bibinfo{volume}{abs/2004.09297}
  (\bibinfo{year}{2020}).
\newblock


\bibitem[Thomson et~al\mbox{.}(2008)]%
        {thomson2008objectivity}
\bibfield{author}{\bibinfo{person}{Elizabeth~A Thomson},
  \bibinfo{person}{Peter~RR White}, {and} \bibinfo{person}{Philip Kitley}.}
  \bibinfo{year}{2008}\natexlab{}.
\newblock \showarticletitle{“Objectivity” and “hard news” reporting
  across cultures: Comparing the news report in English, French, Japanese and
  Indonesian journalism}.
\newblock \bibinfo{journal}{\emph{Journalism studies}} \bibinfo{volume}{9},
  \bibinfo{number}{2} (\bibinfo{year}{2008}), \bibinfo{pages}{212--228}.
\newblock


\bibitem[Vatolin et~al\mbox{.}(2021)]%
        {Vatolin2021-ja}
\bibfield{author}{\bibinfo{person}{A~S Vatolin}, \bibinfo{person}{{SberBank /
  Moscow, Russia}}, \bibinfo{person}{E~Y Smirnova}, \bibinfo{person}{S~S
  Shkarin}, {and} \bibinfo{person}{{SberBank / Moscow, Russia}}.}
  \bibinfo{year}{2021}\natexlab{}.
\newblock \showarticletitle{Russian News Similarity Detection with {SBERT}:
  pre-training and fine-tuning}. \bibinfo{publisher}{Russian State University
  for the Humanities}.
\newblock


\bibitem[Vrijenhoek et~al\mbox{.}(2021)]%
        {Vrijenhoek2021-zh}
\bibfield{author}{\bibinfo{person}{Sanne Vrijenhoek}, \bibinfo{person}{Mesut
  Kaya}, \bibinfo{person}{Nadia Metoui}, \bibinfo{person}{Judith M{\"o}ller},
  \bibinfo{person}{Daan Odijk}, {and} \bibinfo{person}{Natali Helberger}.}
  \bibinfo{year}{2021}\natexlab{}.
\newblock \showarticletitle{Recommenders with a Mission: Assessing Diversity in
  News Recommendations}. In \bibinfo{booktitle}{\emph{CHIIR}} (Canberra ACT,
  Australia) \emph{(\bibinfo{series}{CHIIR '21})}. \bibinfo{pages}{173--183}.
\newblock


\bibitem[Ward et~al\mbox{.}(2013)]%
        {ward2013comparing}
\bibfield{author}{\bibinfo{person}{Michael~D Ward}, \bibinfo{person}{Andreas
  Beger}, \bibinfo{person}{Josh Cutler}, \bibinfo{person}{Matthew Dickenson},
  \bibinfo{person}{Cassy Dorff}, {and} \bibinfo{person}{Ben Radford}.}
  \bibinfo{year}{2013}\natexlab{}.
\newblock \showarticletitle{Comparing GDELT and ICEWS event data}.
\newblock \bibinfo{journal}{\emph{Analysis}} \bibinfo{volume}{21},
  \bibinfo{number}{1} (\bibinfo{year}{2013}), \bibinfo{pages}{267--297}.
\newblock


\bibitem[Weimann and Brosius(2017)]%
        {weimann2017redirecting}
\bibfield{author}{\bibinfo{person}{Gabriel Weimann} {and}
  \bibinfo{person}{Hans-Bernd Brosius}.} \bibinfo{year}{2017}\natexlab{}.
\newblock \showarticletitle{Redirecting the agenda: Agenda-setting in the
  online Era}.
\newblock \bibinfo{journal}{\emph{The Agenda Setting Journal}}
  \bibinfo{volume}{1}, \bibinfo{number}{1} (\bibinfo{year}{2017}),
  \bibinfo{pages}{63--102}.
\newblock


\bibitem[Wilkinson and Thelwall(2012)]%
        {wilkinson2012trending}
\bibfield{author}{\bibinfo{person}{David Wilkinson} {and} \bibinfo{person}{Mike
  Thelwall}.} \bibinfo{year}{2012}\natexlab{}.
\newblock \showarticletitle{Trending Twitter topics in English: An
  international comparison}.
\newblock \bibinfo{journal}{\emph{Journal of the American Society for
  Information Science and Technology}} \bibinfo{volume}{63},
  \bibinfo{number}{8} (\bibinfo{year}{2012}), \bibinfo{pages}{1631--1646}.
\newblock
\urldef\tempurl%
\url{https://doi.org/10.1002/asi.22713}
\showDOI{\tempurl}
\showeprint{https://onlinelibrary.wiley.com/doi/pdf/10.1002/asi.22713}


\bibitem[Wu(1998)]%
        {wu1998systemic}
\bibfield{author}{\bibinfo{person}{Haoming~Denis Wu}.}
  \bibinfo{year}{1998}\natexlab{}.
\newblock \bibinfo{booktitle}{\emph{The systemic determinants of international
  news coverage}}.
\newblock \bibinfo{publisher}{The University of North Carolina at Chapel Hill}.
\newblock


\bibitem[Wu(2000)]%
        {wu2000systemic}
\bibfield{author}{\bibinfo{person}{H~Denis Wu}.}
  \bibinfo{year}{2000}\natexlab{}.
\newblock \showarticletitle{Systemic determinants of international news
  coverage: A comparison of 38 countries}.
\newblock \bibinfo{journal}{\emph{Journal of communication}}
  \bibinfo{volume}{50}, \bibinfo{number}{2} (\bibinfo{year}{2000}),
  \bibinfo{pages}{110--130}.
\newblock


\bibitem[Wu(2003)]%
        {wu2003homogeneity}
\bibfield{author}{\bibinfo{person}{H~Denis Wu}.}
  \bibinfo{year}{2003}\natexlab{}.
\newblock \showarticletitle{Homogeneity around the world? Comparing the
  systemic determinants of international news flow between developed and
  developing countries}.
\newblock \bibinfo{journal}{\emph{Gazette (Leiden, Netherlands)}}
  \bibinfo{volume}{65}, \bibinfo{number}{1} (\bibinfo{year}{2003}),
  \bibinfo{pages}{9--24}.
\newblock


\bibitem[Wu(2007)]%
        {wu2007brave}
\bibfield{author}{\bibinfo{person}{H~Denis Wu}.}
  \bibinfo{year}{2007}\natexlab{}.
\newblock \showarticletitle{A brave new world for international news? Exploring
  the determinants of the coverage of foreign news on US websites}.
\newblock \bibinfo{journal}{\emph{International Communication Gazette}}
  \bibinfo{volume}{69}, \bibinfo{number}{6} (\bibinfo{year}{2007}),
  \bibinfo{pages}{539--551}.
\newblock


\bibitem[Xiao et~al\mbox{.}(2021)]%
        {Xiao2021-zh}
\bibfield{author}{\bibinfo{person}{Kejing Xiao}, \bibinfo{person}{Zhaopeng
  Qian}, {and} \bibinfo{person}{Biao Qin}.} \bibinfo{year}{2021}\natexlab{}.
\newblock \showarticletitle{A graphical decomposition and similarity
  measurement approach for topic detection from online news}.
\newblock \bibinfo{journal}{\emph{Inf. Sci.}}  \bibinfo{volume}{570}
  (\bibinfo{date}{Sept.} \bibinfo{year}{2021}), \bibinfo{pages}{262--277}.
\newblock


\bibitem[Xu et~al\mbox{.}(2019)]%
        {Xu2019-ym}
\bibfield{author}{\bibinfo{person}{Guixian Xu}, \bibinfo{person}{Yueting Meng},
  \bibinfo{person}{Zhan Chen}, \bibinfo{person}{Xiaoyu Qiu},
  \bibinfo{person}{Changzhi Wang}, {and} \bibinfo{person}{Haishen Yao}.}
  \bibinfo{year}{2019}\natexlab{}.
\newblock \showarticletitle{Research on Topic Detection and Tracking for Online
  News Texts}.
\newblock \bibinfo{journal}{\emph{IEEE Access}}  \bibinfo{volume}{7}
  (\bibinfo{year}{2019}), \bibinfo{pages}{58407--58418}.
\newblock


\bibitem[Xu et~al\mbox{.}(2022)]%
        {xu-etal-2022-hfl}
\bibfield{author}{\bibinfo{person}{Zihang Xu}, \bibinfo{person}{Ziqing Yang},
  \bibinfo{person}{Yiming Cui}, {and} \bibinfo{person}{Zhigang Chen}.}
  \bibinfo{year}{2022}\natexlab{}.
\newblock \showarticletitle{{HFL} at {S}em{E}val-2022 Task 8: A
  Linguistics-inspired Regression Model with Data Augmentation for Multilingual
  News Similarity}. In \bibinfo{booktitle}{\emph{SemEval-2022}}.
  \bibinfo{publisher}{Association for Computational Linguistics},
  \bibinfo{address}{Seattle, United States}, \bibinfo{pages}{1114--1120}.
\newblock
\urldef\tempurl%
\url{https://doi.org/10.18653/v1/2022.semeval-1.157}
\showDOI{\tempurl}


\bibitem[Yang et~al\mbox{.}(2020)]%
        {yang-etal-2020-multilingual}
\bibfield{author}{\bibinfo{person}{Yinfei Yang}, \bibinfo{person}{Daniel Cer},
  \bibinfo{person}{Amin Ahmad}, \bibinfo{person}{Mandy Guo},
  \bibinfo{person}{Jax Law}, \bibinfo{person}{Noah Constant},
  \bibinfo{person}{Gustavo Hernandez~Abrego}, \bibinfo{person}{Steve Yuan},
  \bibinfo{person}{Chris Tar}, \bibinfo{person}{Yun-hsuan Sung},
  \bibinfo{person}{Brian Strope}, {and} \bibinfo{person}{Ray Kurzweil}.}
  \bibinfo{year}{2020}\natexlab{}.
\newblock \showarticletitle{Multilingual Universal Sentence Encoder for
  Semantic Retrieval}. In \bibinfo{booktitle}{\emph{ACL: System
  Demonstrations}}. \bibinfo{publisher}{Association for Computational
  Linguistics}, \bibinfo{address}{Online}, \bibinfo{pages}{87--94}.
\newblock
\urldef\tempurl%
\url{https://doi.org/10.18653/v1/2020.acl-demos.12}
\showDOI{\tempurl}


\bibitem[Yang et~al\mbox{.}(2002)]%
        {yang2002topic}
\bibfield{author}{\bibinfo{person}{Yiming Yang}, \bibinfo{person}{Jian Zhang},
  \bibinfo{person}{Jaime Carbonell}, {and} \bibinfo{person}{Chun Jin}.}
  \bibinfo{year}{2002}\natexlab{}.
\newblock \showarticletitle{Topic-conditioned novelty detection}. In
  \bibinfo{booktitle}{\emph{Proceedings of the eighth ACM SIGKDD international
  conference on Knowledge discovery and data mining}}.
  \bibinfo{pages}{688--693}.
\newblock


\bibitem[Yesilbas et~al\mbox{.}(2021)]%
        {yesilbas2021analysis}
\bibfield{author}{\bibinfo{person}{Veysel Yesilbas}, \bibinfo{person}{Jose~J
  Padilla}, {and} \bibinfo{person}{Erika Frydenlund}.}
  \bibinfo{year}{2021}\natexlab{}.
\newblock \showarticletitle{An analysis of global news coverage of refugees
  using a big data Approach}. In \bibinfo{booktitle}{\emph{SBP-BRiMS 2021,
  Proceedings 14}}. Springer, \bibinfo{pages}{111--120}.
\newblock


\bibitem[Zuckerman(2003)]%
        {zuckerman2003global}
\bibfield{author}{\bibinfo{person}{Ethan Zuckerman}.}
  \bibinfo{year}{2003}\natexlab{}.
\newblock \showarticletitle{Global Attention Profiles-A working paper: First
  steps towards a quantitative approach to the study of media attention}.
\newblock \bibinfo{journal}{\emph{Berkman Center Research Publication}}
  \bibinfo{number}{2003-06} (\bibinfo{year}{2003}).
\newblock


\bibitem[Zuckerman(2013)]%
        {zuckerman2013rewire}
\bibfield{author}{\bibinfo{person}{Ethan Zuckerman}.}
  \bibinfo{year}{2013}\natexlab{}.
\newblock \bibinfo{booktitle}{\emph{Rewire: Digital cosmopolitans in the age of
  connection}}.
\newblock \bibinfo{publisher}{WW Norton \& Company}.
\newblock


\end{thebibliography}

\appendix

\section*{Appendix A: Transformer model of news similarity}

\textbf{News article input selection. }
News articles can be quite long, yet most of the state-of-the-art deep learning models have a restricted input length, e.g., 512 word-pieces \cite{reimers-2019-sentence-bert, Song2020MPNetMA, yang-etal-2020-multilingual}. Therefore, we have to select a subset of the text of each news article to embed. Given that most articles follow the pyramid writing format \cite{po2003news, thomson2008objectivity}, the important information is often at the start; hence, we select part of the head and a part of the tail of a news. We treat decision on how much of the head vs. the tail to select as a hyper-parameter we optimize for.

\textbf{News similarity model architectures. }
There are two types of mainstream encoders for text similarity: (1) cross-encoders which accept two texts as inputs and output a similarity score directly;  and (2) bi-encoders which embed each input document separately as a fixed-length vector, and the similarity is then computed as the distance between two embeddings. We use a bi-encoder structure for two reasons. First, bi-encoders are more computationally efficient: the document representations for the bi-encoder may be pre-computed once and then used in all comparisons, while a cross-encoder requires computing the representations of each pair separately, thus incurring an $O(n^2)$ cost which is computational infeasible for millions of articles in our data. Second, because each document is encoded individually, bi-encoders allow for more text to included as input and thus capture more information. Top performing systems in the 2022 SemEval task on news similarity \cite{chen2022semeval} and recent work developing these models for scale \cite{litterer2023rains} suggest that bi-encoders can achieve comparable performance to cross-encoders, which is also mirrored our experimentation.

\textbf{News similarity mapping. }
The news article similarity was labeled using a four point scale [1,4] for Very Dissimilar ($x=1$), Somewhat Dissimilar ($x=2$), Somewhat Similar ($x=3$), and Very Similar ($x=4$). 
We use labels of each annotator individually to train the news similarity models.

There are multiple ways to map these values into the same range as cosine similarity to compute the training loss. 
We can map labels from the four point scale [1,4] to range $[l, r]$ via a linear transformation $t=px+q$ without the loss of functional monotonicity. In this case two boundary conditions to news are $l=p+q$ and $r=4p+q$ so that $p=\frac{(r-l)}{3}$ and $q=\frac{(-r+4l)}{3}$. Specifically we attempted an ``unsigned" transformation $t_u$ where $l_u=0$ and $r_u=1$, as well as a ``signed" transformation $t_s$ where $l_s=-1$ and $r_s=1$ when we optimize our model performance. 

In addition, we observed that annotators spent much more time on distinguishing Somewhat Similar or Somewhat Dissimilar pairs than identifying Very Similar and Very Dissimilar pairs, which indicates the decision boundary is vague across ``Similar" and ``Dissimilar." We therefore experiment with transforming the optimization objective to make its learning gradient steeper around this decision boundary, so that the model can better learn the decision boundary. 

To this end, we use cubed unsigned and signed similarity values, $\phi=t_u^3$ and $\phi=\frac{(2t_s-1)^3}{2} + \frac{1}{2}$, respectively, where $\phi$ is the output of: (1) cross-encoders or (2) the result of cosine similarity between two news article embeddings in the case of bi-encoders.


\textbf{Multi-label learning. }
In our data annotation, the news similarity is ordinal, but in realistic scenario the similarity should be a continuous, real-valued function. So annotation causes inevitably precision loss of similarity even though the annotators have a very good understanding of news similarity in their minds. We posit that the similarity ratings of other aspects (other than \textsc{Overall}) may also reveal information as to overall similarity of a news pair. By introducing them into the learning objective, precision loss from annotation can be mitigated. Specifically we devised two integrated similarities $y_1$ and $y_2$ for training, which incorporate the average of multiple aspects for a single pair:

\begin{equation}
    y_1 = w_1 * x + (1 - w_1) * \overline{x_i},
\end{equation}
where $x_i$ is the similarity of the news pair in dimension $i$, $y$ is the overall similarity of the news pair, $i\in$ \{ENT, NAR\}, and

\begin{equation}
\begin{aligned}
    y_2 & =  w_2 * x + (1 - w_2) * \overline{x_j} &
\end{aligned}
\end{equation}
where $j\in$ \{GEO, ENT, TIME, NAR, STYLE, TONE\}. The weights $w_1$ and $w_2$ can be optimized by empirical study of model performance. During the training process, the learning objective of the model is to minimize the difference from its output to the integrated similarity, i.e. min $|y_i-\phi|$ where $i\in\{1, 2\}$. 

\textbf{Experiments.}
We optimized the selection of the proposed approaches and corresponding hyper-parameters according to model performance. In this way we identify the best model for our news article similarity task.

We take 64\% of our dataset for training, 16\% for development, and 20\% for testing. We evaluated multiple large pre-trained multilingual deep learning models as the base of comparison; these models  are the state-of-the-art for textual similarity: LaBSE~\cite{feng-etal-2022-language}, MPNet~\cite{Song2020MPNetMA}, and MUSE~\cite{yang-etal-2020-multilingual}. Also we compared two state-of-the-art models in the specific news comparison task: HFL \cite{xu-etal-2022-hfl} and Gatenlp\cite{singh-etal-2022-gatenlp}. Table \ref{tab:model_performance} shows the summary. All models are trained on 4x NVIDIA Tesla M40,
and the news similarities at global scale are inferred on 8x NVIDIA GeForce GTX 2080 Ti. 


LaBSE outperformed the other base models, perhaps supported by its rich multilingual pre-training data. In general, all base models perform much worse than the fine-tuned models, which reveals the uniqueness of the news similarity task---i.e., the task is not just measuring textual similarity---and thus the need of developing a dataset for this task. By empirical test, we found the best hyperparameter configuration of our multiligual news similarity model is to take the first 456 wordpieces and last 56 wordpieces from each news article, with the basis of LaBSE model. The best performance was found using the ``unsigned” range for embedding space with $t_u$ and use $y_1$ as the integrated similarity for learning with $w_1=0.8$. Our best bi-encoder model is able to achieve comparable performance to the HFL cross-encoder model while being an order of magnitude times faster.

\section*{Appendix B: News article pair filtering}

\begin{figure}[t!]
\centering
\includegraphics[width=0.45\columnwidth]{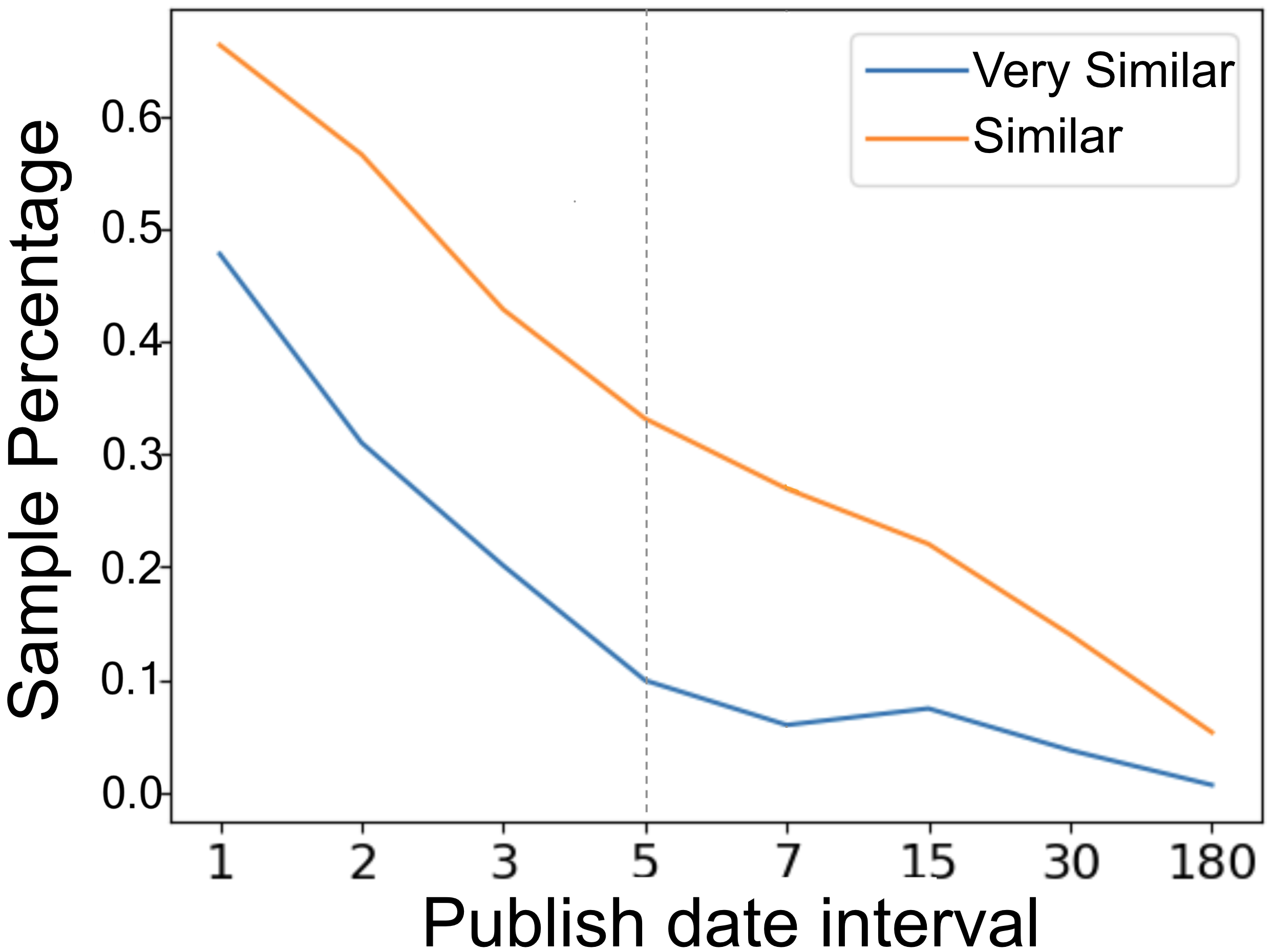}
\hspace{4mm}
\includegraphics[width=0.47\columnwidth]{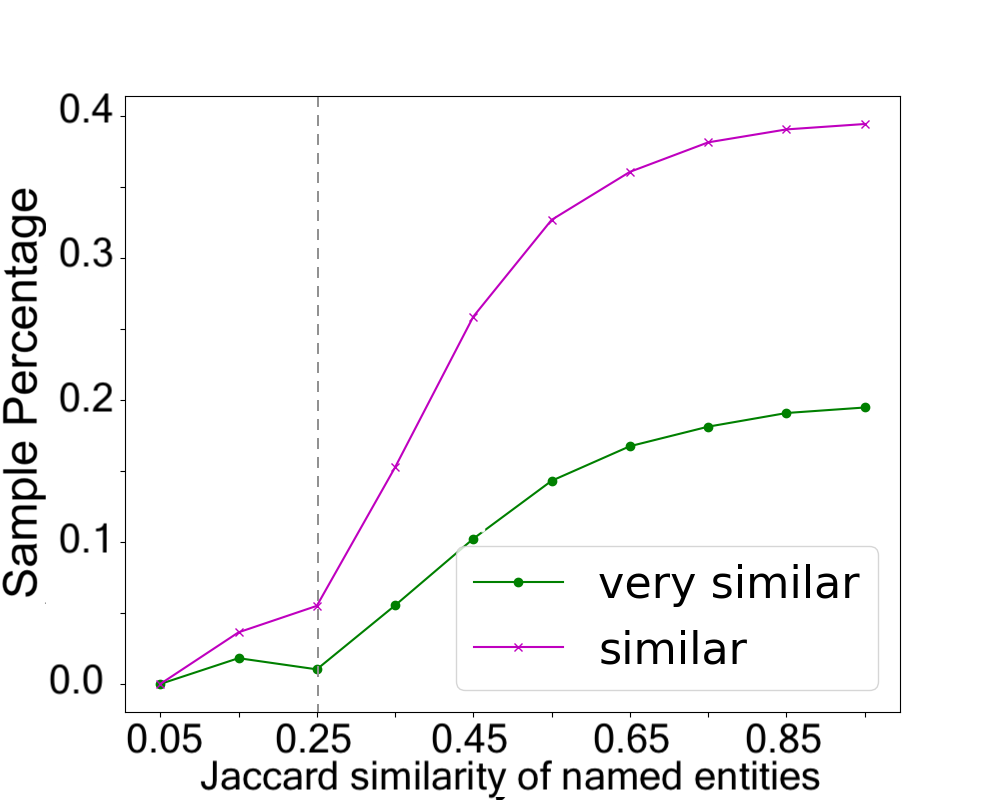}
\caption{Distribution of publish date intervals (left) and Jaccard similarity of named entities (right) for the news article pairs labeled as (somewhat) similar by human annotators in the extended dataset of \citet{chen2022semeval}.}
\label{distrubtion_sample}
\end{figure}

In Section \ref{sec:news-net}, we applied a set of thresholds to select the news article pairs for annotation. 
We choose the thresholds by aiming for a high fraction of similar news article pairs among all annotated pairs (Figure \ref{distrubtion_sample}).
In this way, we selected the news article pairs that are published within a time frame not exceeding five days apart, coupled with a Jaccard similarity coefficient exceeding 0.25 for named entities. The sample percentage represents the percentage of news article pairs that are labeled as (somewhat or very) similar in all the pairs under the same condition (publish date interval or Jaccard similarity of named entities).

\section*{Appendix C: Similarity of news between and within news event clusters.}

As a supplementary evaluation of the quality of the news clusters, we briefly discuss their face validity and robustness. To assess face validity, we analyzed ground-truth human assessment of news similarity within and between clusters (as shared in the news similarity dataset by \citet{chen2022semeval}). 

\begin{figure}[t!]
\centering
\includegraphics[width=0.88\columnwidth]{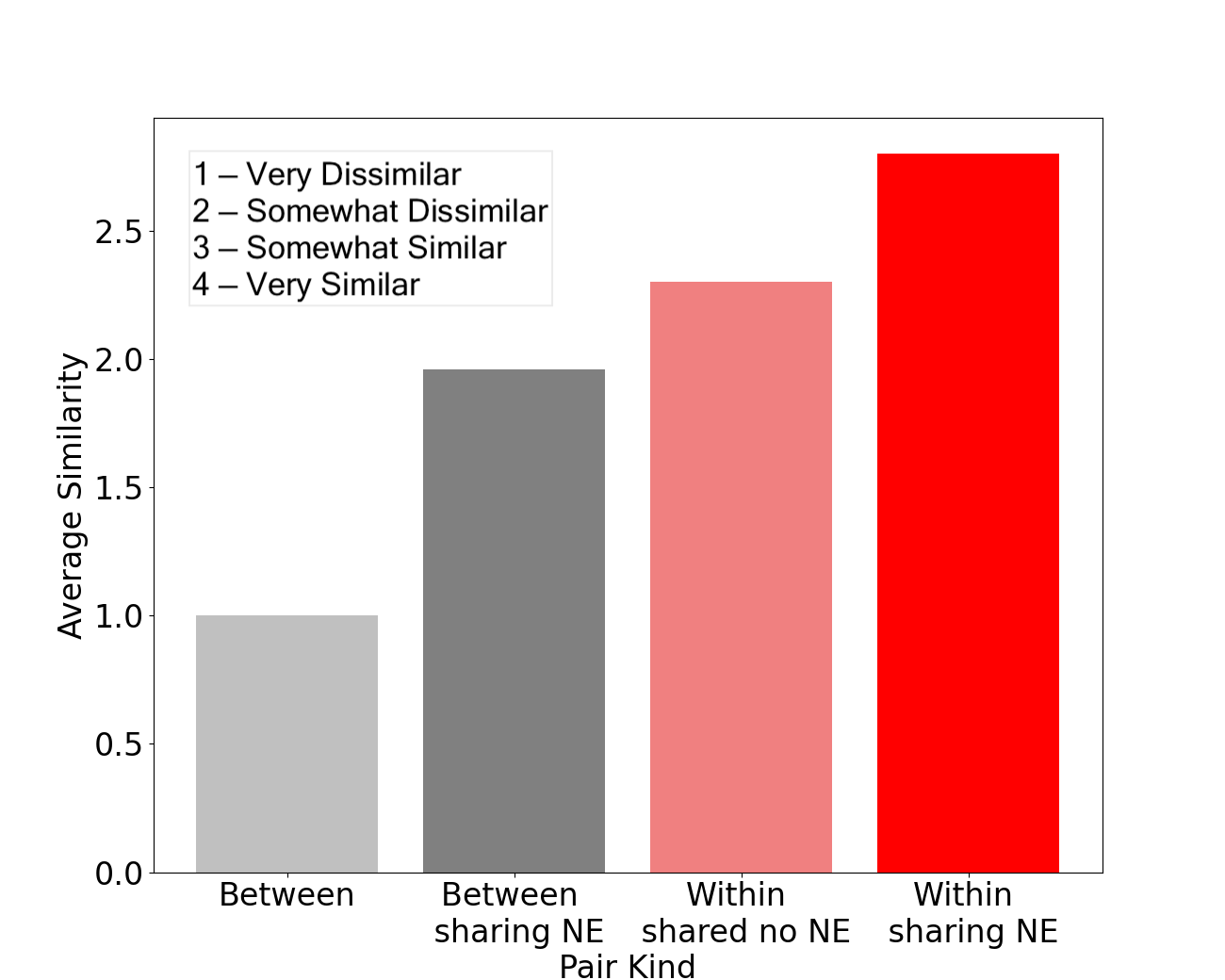}
\caption{Average similarity of news article pairs between and within clusters, sharing or not sharing named entities (NE), using the similarity scale of \citet{chen2022semeval}.}
\label{fig:aver_sim_cmp}
\end{figure}

As Figure \ref{fig:aver_sim_cmp} presents, pairs of news articles residing within the same event cluster, irrespective of whether they share any named entities, demonstrate a higher degree of similarity compared to article pairs from disparate clusters, even those that do share named entities. This suggest that sharing a event cluster is a more potent indicator of news article similarity than sharing a named entity.

\section*{Appendix D: Predictors used to explain NEWS COVERAGE ACROSS COUNTRIES}

We surveyed extensive literature for traits of countries and international relations that are related to similar news practices, and categorized them into five areas: economic, political, societal, geographic and cultural \cite{lee2007international, segev2016group}. Explanation for each area are presented in the rest of this section. Each factor is either numeric or categorical, as described in Table \ref{tab:country_factors_details} (included only in the ArXiv version of this manuscript).

\textbf{Economic.} Economy has been widely found as a strong factor to influence media content and news flow in prior studies \cite{lee2007international, segev2016group, wu2000systemic}. On the one side, financial limitations of media outlets, particularly their dependence on advertising revenue, play a major role in shaping their editorial policies \cite{mccombs1993evolution}. A typical indicator of national economy level can be GDP \cite{segev2015visible, guo2017global}. On the other side, modern news organizations become market-oriented and thus are often more focused on covering stories that will generate the most clicks, views, or ad revenue. For example, a drop of United States in Chinese trade volume would also lead to a decrease in its attraction in Chinese news coverage and its agenda-setting power on Chinese media. Also low income countries usually have different varying needs for international news~\cite{wu2003homogeneity}.



Specific economic factors include a national trait: GDP~\cite{lee2007international, segev2015visible, segev2016group, grasland2020international, guo2020predictors}; and two international relatedness measures: trade \cite{lee2007international, kariel1984factors, segev2016group, wu2007brave, dupree1971international, guo2020predictors} and investment volume \cite{lee2007international}.

\textbf{Political.} The governments can manipulate journalistic output with their regulatory and legislative powers. Also the reporters rely on dominant social and political institutions for routine access to a significant volume of newsworthy information so that their deadlines and demands can be met \cite{greer2003sex}. Therefore how the countries distribute and execute political power internally or externally also play a role in shaping the perspectives and agendas of news coverage. Furthermore military also affects the media agenda.

Specific political factors include national traits and international relatedness in terms of political freedom (democracy index) \cite{segev2016group, kim1996determinants, nnaemeka1981internal, robinson1976international}, press freedom \cite{wu1998systemic, wu2007brave}, government mode (republic and federalism) \cite{hester1973theoretical, wilkinson2012trending}, diplomatic \cite{lee2007international, guo2017global, guo2020predictors}, military power (military strength index)~\cite{wu2007brave, segev2015visible, kim1996determinants, guo2020predictors}, and conflict intensity (global peace index)  \cite{segev2015visible, segev2016group}.

\textbf{Societal.} As the foundation of society, people naturally become a indispensable part in the media system. When news are spread, not only the editors and journalists decide news coverage policy \cite{greer2003sex}, but also the audiences' attention form the media market and thus affect the newsworthiness \cite{weimann2017redirecting}. Population size and immigration size are two typical indicators of people diversity, and population density can impact the frequency and strength of social connections between people. Social inequality contribute audience diversity too as individuals from different socioeconomic classes may have different interests\cite{alexander2018digital,bourdieu1993field}, which can be reflected by gini index.



Specifically societal factors include population \cite{lee2007international, kariel1984factors, segev2016group, grasland2020international, kim1996determinants, dupree1971international}, immigration\cite{wu2007brave, segev2015visible, segev2016group, jansen2019drives}, density \cite{dupree1971international}, and gini index \cite{segev2016group}.

\textbf{Geographic.} Geographic size and location of a country determines the pattern of its internal and external communications and thus shape its news coverage. When events occur far away from the media outlet's base, reporters may face logistical and financial challenges in covering them, which limit the amount and quality of coverage. Additionally, distance can create language and cultural barriers that may make it harder for journalists to fully understand and accurately report on events. 


Specifically geographic factors include area \cite{lee2007international, segev2016group, grasland2020international}, distance \cite{kariel1984factors, segev2016group, segev2013america, grasland2020international, kim1996determinants, dupree1971international}, shared border \cite{grasland2020international, kariel1984factors, segev2015visible, wu2007brave, golan2016can, guo2020predictors}, and shared continent or region \cite{kim1996determinants, dupree1971international, guo2020predictors}

\textbf{Cultural.} One way that culture shape media agendas is through the advocated values and norms within a particular society. For example, some cultures have a strong emphasis on community and social responsibility, which may lead media outlets to prioritize stories about local issues and social justice. While some other cultures may have a greater emphasis on individualism and consumerism, which may result in more media attention to lifestyle and entertainment. Culture can also influence the way that news stories are framed and interpreted by media outlets, depending on factors such as language, symbolism, and historical context.


Specifically cultural factors include language \cite{kariel1984factors, segev2016group, segev2013america, wu2007brave, kim1996determinants, dupree1971international}, religion \cite{lee2007international, segev2016group, segev2013america}, literacy rate \cite{lee2007international, segev2016group, dupree1971international}, Internet user rate (does not appear in the literature), and media outlet numbers \cite{wu2003homogeneity, wu2007brave, segev2016group, gerbner1977many}.



\begin{table*}[t!]
\resizebox{2\columnwidth}{!}{
    \NoHyper
    \begin{tabular}{r p{0.7\columnwidth} p{0.95\columnwidth}}

    \toprule
    Type & Predictor & Type\\
    \toprule
   \multirow{3}{*}{Econ.} & \text{{\color{orange} Trade}}  \cite{lee2007international, kariel1984factors, segev2016group, wu2007brave, dupree1971international, guo2020predictors} & numeric\\    
            & \text{{\color{orange} GDP}}  \cite{lee2007international, segev2015visible, segev2016group, grasland2020international, guo2020predictors} & numeric\\
            & \text{{\color{orange} Financial Investment}}  \cite{lee2007international} & numeric \\
    \hline
    \multirow{7}{*}{Pol.} & \text{{\color{blue} Democracy index} }  \cite{ robinson1976international, nnaemeka1981internal, kim1996determinants, segev2016group} & numeric\\
                         & \text{{\color{blue} Political blocs}} & categorical from \{NATO, EU, BRICS\} \\
                         & \text{{\color{blue} Press freedom}}  \cite{wu1998systemic, wu2007brave} & numeric\\ 
                         &  \text{{\color{blue} Mode of government}}  \cite{hester1973theoretical, wilkinson2012trending} & categorical from \{Federalism, Unitary republic, Other\} \\ 
                         & \text{{\color{blue} Diplomatic ties}}  \cite{lee2007international, guo2017global, guo2020predictors} & categorical from \{Ambassador Relation, Other\} \\
                         & \text{{\color{blue} Military strength index}}  \cite{wu2007brave, segev2015visible, kim1996determinants, guo2020predictors} & numeric\\
                         & \text{{\color{blue} Conflict intensity (global peace index)}}  \cite{segev2015visible, segev2016group} & numeric\\
    \hline
    \multirow{4}{*}{Soc.} &  \text{{\color{brown} Population}}   \cite{lee2007international, kariel1984factors, segev2016group, grasland2020international, kim1996determinants, dupree1971international} & numeric\\
                        & \text{{\color{brown} Immigration}}   \cite{wu2007brave, segev2015visible, segev2016group, jansen2019drives} & numeric\\
                        & \text{{\color{brown} Population density}}   \cite{dupree1971international} & numeric\\ 
                        & \text{{\color{brown} Gini index}}   \cite{segev2016group} & numeric \\
    \hline
    \multirow{4}{*}{Geo.} & \text{{\color{green} Distance}}  \cite{kariel1984factors, segev2016group, segev2013america, grasland2020international, kim1996determinants, dupree1971international} & numeric\\
                          & \text{{\color{green} Neighbors}}   \cite{grasland2020international, kariel1984factors, segev2015visible, wu2007brave, guo2020predictors} & categorical from \{Neighbors, Other\}\\
                          & \text{{\color{green} Area}}  \cite{lee2007international, segev2016group, grasland2020international} & numeric \\
                          & \text{{\color{green} Continent}}  \cite{kim1996determinants, dupree1971international, guo2020predictors} & categorical from \{Asia, Africa, North America, South America, Antarctica, Europe, and Australia\} \\
    \hline
    \multirow{5}{*}{Cul.} & \text{{\color{purple} Language}}  \cite{kariel1984factors, segev2016group, segev2013america, wu2007brave, kim1996determinants, dupree1971international} & categorical from \{English, Spanish, German, French, Chinese, Polish, Turkish, Italian, Arabic, Russian\}\\
                        & \text{{\color{purple} Religion}}  \cite{lee2007international, segev2016group, segev2013america} & categorical from \{Christian, Muslim, Unaffil, Hindu,	Buddhist, Jewish, Folk religion, Other religion\}\\
                        & \text{{\color{purple} Media outlets number}}  \cite{wu2003homogeneity, wu2007brave, segev2016group, gerbner1977many} & numeric \\
                        & \text{{\color{purple} Literacy rate}}  \cite{lee2007international, segev2016group, dupree1971international} & numeric\\ 
                        & \text{{\color{purple} Internet user rate}} & numeric \\
    \bottomrule
    \end{tabular}
    }
    \caption{Predictor types considered in the within-country and between-country models with references to studies that mention them. }
    \label{tab:country_factors_details}
\end{table*}

\section*{Appendix E: BASELINE NEWS SYNCHRONY MEASURE}

We mirrored the analysis in \S\ref{sec:multivariate-regression} using the pairwise news similarity and obtained qualitatively similar results, as shown in Table \ref{tab:intra_country_details} and Table \ref{tab:inter_country_details} (blank values indicate the factors are not excluded from the model by the feature selection process).

\section*{Appendix F: Robustness Check for Regression analysis}

In order to check the robustness of our results in the regression analysis, we also applied Gradient Boosting Regression Trees (GBRT) on the news diversity and synchrony. Table \ref{tab:inter_country_details_tree} shows the feature importance of news synchrony regression, yielding an adjusted R-squared value of 0.60. Notably, trade volume and the presence of a shared language remain the two factors with the most importance. However, it's important to note that when applying GBRT to the news diversity, we observed a tendency for the model to overfit, given there are only a little more than 100 aggregated samples of countries. Therefore we omit its result in this paper.

\begin{table*}

\resizebox{0.55\textwidth}{!}{
    \begin{tabular}{ccccccccc}
    \toprule
    Predictor & \makecell{VIF\\Cluster\\similarity\\model\\coeeficient} & \makecell{VIF\\Cluster\\similarity\\model\\p-value} &  \makecell{AIC\\Cluster\\similarity\\model\\coeeficient} & \makecell{AIC\\Cluster\\similarity\\model\\p-value} \\
    \midrule
    
    {\color{gray} Intercept} & \ 0.653 & 0.003 &  \ 0.586 & 0.005 \\
    
    {\color{blue} Republic?} & & & &  \\
    {\color{blue} Federalism?} & \textbf{-0.118} & \textbf{0.002} & \textbf{-0.069} & \textbf{0.002} \\

    {\color{blue} Within NATO?} & -0.726 & 0.002 & -0.044 & 0.002    \\
    {\color{blue} Within EU?} & \ 0.028 & 0.004 & \ 0.093 & 0.007   \\
    \text{{\color{blue} Within BRICS?}} & -0.006 & 0.003 & -0.012 &0.003  \\
    {\color{blue} Democracy index} &  & & -0.175 & 0.004  \\

    {\color{orange} GDP} & & & &   \\
    {\color{orange} Gini index} & -0.174 & 0.003 & -0.196 & 0.003  \\
    
    {\color{brown} Population} & -0.146 & 0.007 & -0.102 & 0.010    \\
    {\color{brown} Net Migration} &  &  & -0.027 & 0.006 \\
    {\color{brown} Area} & & & -0.093 & 0.007 \\
    
    {\color{pink} Peace index} &  & & \ 0.195 & 0.004  \\
    
    {\color{purple} \#media} & \ 0.148 & 0.004 & \ 0.130 & 0.004    \\
    {\color{purple} Same media?} &  \ 0.069 & 0.001 & \ 0.063 & 0.001  \\
    
    {\color{purple} \#languages} & -0.212 & 0.000 & -0.025 & 0.000   \\
    {\color{purple} \#language families} & -0.058 & 0.004 & -0.038 & 0.004  \\
    {\color{purple} Same language?} & & & &  \\
    {\color{purple} Internet user rate} & -0.300 & 0.003 & -0.371 & 0.004 \\
    {\color{purple} Literacy Rate} &  &  & \ 0.126 & 0.004 \\
    {\color{purple} Religion diversity} & -0.154 & 0.002 & -0.126 & 0.002 \\

    \bottomrule
    \end{tabular}
    }
    \caption{Predictors of within-country similarity regression model based on Ordinary least squares. Negative coefficients mean less similarity and more diversity. Predictor color corresponds to its category. The coefficients align across Cluster model and Similarity model are bold. The strongest predictor is highlighted. Missing value indicates either this factor is dropped during factor selection process. }
    \label{tab:intra_country_details}

\end{table*}

\begin{table*}[t!]
\resizebox{2\columnwidth}{!}{
    \centering

    \begin{tabular}{ccccccccc}
    \toprule
    Predictor & \makecell{VIF\\Cluster\\model\\coefficient} & \makecell{VIF\\Cluster\\model\\p-value} & \makecell{VIF\\Pairwise\\model\\coefficient} & \makecell{VIF\\Pairwise\\model\\p-value} & \makecell{AIC\\Cluster\\model\\coefficient} & \makecell{AIC\\Cluster\\model\\p-value} & \makecell{AIC\\Pairwise\\model\\coefficient} & \makecell{AIC\\Pairwise\\model\\p-value}\\
    \midrule
    {\color{gray} Intercept} & \ 0.2378  & 0.000  & \ 0.6269 & 0.000 & \ 0.2480 & 0.000 & \ 0.6448& 0.000\\
    {\color{green} Continent sim} & \ 0.0385 & 0.000 & -0.0303& 0.000 & \ 0.0372& 0.000 & -0.0305& 0.000\\
    {\color{green} Geographic distance} & & & -0.0389& 0.000 & & & -0.0393& 0.000 \\

    {\color{blue} Same government mode type} & \ 0.0153 & 0.000 & \ 0.0040& 0.281 & \ 0.0128& 0.000 & & \\
    {\color{blue} Both republic} & & & &  & &  & &  \\
    {\color{blue} Both federalism} & \ \textbf{0.0523} & \textbf{0.000} & \ \textbf{0.0100}& \textbf{0.220} & \ \textbf{0.0527}& \textbf{0.000} & \ \textbf{0.0123}& \textbf{0.110}\\
    {\color{blue} Neither republic nor federalism} & & & &  & &  & &  \\
    {\color{blue} Republic and federalism} & & & &  & &  & &  \\
    {\color{blue} Republic and other} & & & & & &  & & \\
    {\color{blue} Federalism and other} & \ 0.0034 & 0.384 & \ 0.0074 & 0.037 & &  & \ 0.0055& 0.053\\

    {\color{blue} Same political bloc} & -0.0319 & 0.012  & &  & -0.0287& 0.000 & & \\
    {\color{blue} Both in NATO} & \ 0.0046 & 0.762 & -0.0082& 0.659 & &  & -0.0235& 0.005\\
    {\color{blue} Both in EU} & & & \ 0.0189& 0.324 & &  & & \\
    {\color{blue} Both in BRICS} & -0.0031& 0.930 & -0.0374& 0.221 & &  & & \\
    
    {\color{blue} Across NATO--EU groups} & & & -0.0189 & 0.273  & &  & & \\
    {\color{blue} Across NATO--BRICS groups} &\  0.0043 & 0.819 & -0.0090& 0.602 & &  & & \\
    {\color{blue} Across EU--BRICS groups} &\  0.0024& 0.891 & \ 0.0096& 0.555 & & & &  \\

    {\color{blue} Same Democracy index category} & & & &  & &  & -0.0156 & 0.005 \\
    {\color{blue} Democracy index (low--low)} & & & &  & &   & & \\
    {\color{blue} Democracy index (high--high)} &\  \textbf{0.0632} & \textbf{0.000} & \ \textbf{0.0375}& \textbf{0.000} & \ \textbf{0.0557} & \textbf{0.000} & \ \textbf{0.0376}& \textbf{0.000} \\
    {\color{blue} Democracy index (low--high)} & \ 0.0082& 0.181 & \ 0.0158& 0.005 & &  & & \\

    {\color{orange} Same GDP category} & & &-0.0008 &  0.803  & &  & & \\
    {\color{orange} GDP (low--low)} & & & &  & &  & & \\
    {\color{orange} GDP (high--high)} &\  0.0420& 0.000 & \ 0.0077 & 0.257 & \ 0.0426& 0.000 & & \\
    {\color{orange} GDP (low--high)} &\  0.0008& 0.813 & &  & &   & & \\

    {\color{purple} Same language} & \ 0.1070 & 0.000 & & & \ 0.1091 & 0.000  & & \\
    {\color{purple} Speak English} & -0.0762 & 0.000 & \ 0.0294& 0.000 & -0.0777& 0.000 & \ 0.0289& 0.000\\
    {\color{purple} Speak German} & \ \textbf{0.0657}& \textbf{0.008}& \ \textbf{0.0702} & \textbf{0.002} & \ \textbf{0.0646}& \textbf{0.008} & \ \textbf{0.0705}& \textbf{0.001}\\
    {\color{purple} Speak Spanish} &  & & -0.0524 & 0.000  & &  & -0.0528& 0.000\\
    {\color{purple} Speak Polish} & & & &   & &  & & \\
    {\color{purple} Speak French} & -0.0426 & 0.001 & \ 0.0556& 0.000 & -0.0436& 0.000 & \ 0.0560& 0.000\\
    {\color{purple} Speak Chinese} & -0.1188 & 0.109 & -0.0305& 0.652 & -0.1187 & 0.108 & & \\
    {\color{purple} Speak Arabic} & \ 0.0138 & 0.243 & -0.0387& 0.000 & &  & -0.0397& 0.000\\
    {\color{purple} Speak Turkish} & & & &  & &  & &  \\
    {\color{purple} Speak Italian} & -0.0855& 0.251 & \ 0.0110& 0.871 & &  & & \\
    {\color{purple} Speak Russian} &  & & &  & &  & &  \\

    \bottomrule
    \end{tabular}
    }
    
    \caption{Predictors of between-country similarity where country pairs are taken as samples. Negative coefficients mean less similarity and more diversity (with country pairs). Predictor color corresponds to its category. The coefficients align across Cluster model and Similarity model are bold. The strongest predictor is highlighted. Missing value indicates this factor is dropped during factor selection process. GDP is binarized as ``high" (\textgreater \$500 billion) and ``low" (\textless \$500 billion). Democracy index  is encoded as ``high" (\textgreater 5) and ``low" (\textless 5)}
    \label{tab:inter_country_details}
\end{table*}

\begin{table*}[t!]
\resizebox{0.7\columnwidth}{!}{
    \centering

    \begin{tabular}{cc}
    \toprule
    Predictor & \makecell{VIF\\Cluster\\model\\feature\\importance}\\
    \midrule
    {\color{green} Continent sim} & 0.0124\\
    {\color{green} Neighbors} & 0.0054\\

    {\color{blue} Same government mode type} & 0.0023\\
    {\color{blue} Both federalism} & 0.0392\\
    {\color{blue} Federalism and other} & 0.0120\\

    {\color{blue} Both in NATO} & 0.0010\\
    {\color{blue} Both in EU} & 0.0032\\
    {\color{blue} Both in BRICS} & 0.0058\\
    
    {\color{blue} Across NATO--BRICS groups} & 0.0095\\
    {\color{blue} Across EU--BRICS groups} & 0.0046\\

    {\color{blue} Democracy index (high--high)} & 0.1012\\
    {\color{blue} Democracy index (low--high)} & 0.0210\\

    \textbf{{\color{orange} Trade}} & \textbf{0.2887}\\
    {\color{orange} Same GDP category} & 0.0154\\
    {\color{orange} Gini Index(high--high)} & 0.0485\\
    {\color{orange} Gini Index(low--high)} & 0.0130\\

    \textbf{{\color{purple} Same language}} & \textbf{0.1653}\\
    {\color{purple} Speak English} & 0.1003\\
    {\color{purple} Speak French} & 0.0030\\
    {\color{purple} Speak Arabic} & 0.0261 \\
    {\color{purple} Speak Italian} & 0.0015\\

    \bottomrule
    \end{tabular}
    }
    
    \caption{Feature importance of news synchrony regression with gradient boosting trees where country pairs are taken as samples. Negative coefficients mean less similarity and more diversity (with country pairs). Predictor color corresponds to its category. The coefficients align across Cluster model and Similarity model are bold. The strongest predictor is highlighted. Missing value indicates this factor is dropped during factor selection process. GDP is binarized as ``high" (\textgreater \$500 billion) and ``low" (\textless \$500 billion). Democracy index  is encoded as ``high" (\textgreater 5) and ``low" (\textless 5)}
    \label{tab:inter_country_details_tree}
\end{table*}

\end{document}